\documentclass[preprint,12pt]{elsarticle}

\usepackage{amssymb}
\usepackage{xcolor}
\usepackage{soul}
\usepackage[labelformat=simple]{subcaption}

\usepackage{tikz}
\usepackage{pgfplots}
\usepackage{amsmath}
\usepackage{hyperref}
\usetikzlibrary{decorations.pathreplacing}

\usepackage{siunitx}
\definecolor{tomato}{HTML}{FF6347}

\newcommand{\hypar}{\textsc{HyPar}}

\pgfplotsset{
    legend style={font=\small}
}

\makeatletter
\def\ps@pprintTitle{%
 \let\@oddhead\@empty
 \let\@evenhead\@empty
 \def\@oddfoot{\centerline{\thepage}}%
 \let\@evenfoot\@oddfoot}
\makeatother

\begin{document}

\begin{frontmatter}

\title{GPU-Accelerated DNS of Compressible Turbulent Flows\tnoteref{doe}\tnoteref{accepted}}
\tnotetext[doe]{This research was supported by the Exascale Computing Project (17-SC-20-SC), a collaborative effort of the U.S. Department of Energy Office of Science and the National Nuclear Security Administration and by the U.S. Department of Energy, Office of Science, under contract number DE-AC02-06CH11357. Y. Kim and R. Balakrishnan were partially supported by the ECP/ProxyApps sub-project. E. Constantinescu was partially supported by U.S. Department of Energy, Office of Science, Office of Advanced Scientific Computing Research (ASCR). Part of this work was performed under the auspices of the U.S. Department of Energy by Lawrence Livermore National Laboratory under Contract No. DE-AC52-07NA27344.}
\tnotetext[accepted]{This work has been accepted to Computers and Fluids Journal for publication: \url{https://doi.org/10.1016/j.compfluid.2022.105744}.}

\author[MCS]{Youngdae Kim\fnref{youngdae}}
\fntext[youngdae]{Present address: ExxonMobil Technology and Engineering Company, Annandale, NJ 08801, United States. This work was done while the author was at Argonne National Laboratory as a postdoctoral appointee.}
\ead{youngdaekim26@gmail.com}
\author[LLNL]{Debojyoti Ghosh\corref{debo}}
\cortext[debo]{Corresponding author}
\ead{ghosh5@llnl.gov}
\author[MCS]{Emil M. Constantinescu}
\ead{emconsta@anl.gov}
\author[CPS]{Ramesh Balakrishnan}
\ead{bramesh@anl.gov}

\affiliation[CPS]{organization={Computational Science Division, Argonne National Laboratory},
            addressline={9700 S. Cass Ave.},
            city={Lemont},
            postcode={60439},
            state={IL},
            country={United States}}
\affiliation[LLNL]{organization={Center for Applied Scientific Computing, Lawrence Livermore National Laboratory},
            addressline={7000 East Ave.},
            city={Livermore},
            postcode={94550},
            state={CA},
            country={United States}}
\affiliation[MCS]{organization={Mathematics and Computer Science Division, Argonne National Laboratory},
            addressline={9700 S. Cass Ave.},
            city={Lemont},
            postcode={60439},
            state={IL},
            country={United States}}

\begin{abstract}
This paper explores strategies to transform an existing CPU-based high-performance computational fluid dynamics  solver, \hypar{}, for compressible flow simulations on emerging exascale heterogeneous (CPU+GPU) computing platforms. The scientific motivation for developing a GPU-enhanced version of \hypar{} is to simulate canonical turbulent flows at the highest resolution possible on such platforms. We show that optimizing memory operations and thread blocks results in $200$x speedup of computationally intensive kernels compared with a CPU core. Using multiple GPUs and CUDA-aware MPI communication, we demonstrate both strong and weak scaling of our GPU-based \hypar{} implementation on the NVIDIA Volta V100 GPUs. We simulate the decay of homogeneous isotropic turbulence in a triply periodic box on grids with up to $1024^3$ points ($5.3$ billion degrees of freedom) and on up to $1,024$ GPUs. We compare the wall times for CPU-only and CPU+GPU simulations. The results presented in the paper are obtained on the Summit and Lassen supercomputers at  Oak Ridge and Lawrence Livermore National Laboratories, respectively.
\end{abstract}


\end{frontmatter}


\section{Introduction}
\label{sec:intro}
Turbulence is characterized by the seemingly chaotic, yet correlated, motion of fluid elements (in both the Eulerian and Lagrangian sense) over a range of spatial and temporal scales. The distinguishing features of turbulence, however, are the lack of clear separation between the scales of motion (unlike in the kinetic theory of gases), the large number of degrees of freedom, and the large range of length and time scales. Although it is accepted that the dynamics of turbulence is described by the Navier--Stokes equations~(NSE), it is prohibitively expensive to simulate all the length and time scales of motion. For many cases of interest, it is practically impossible to simulate turbulence without appropriate models to account for the effect of the small scales on the evolution of the turbulent flow field. As a result, various theoretical statistical models have been proposed to explain the underlying mechanisms in turbulent flow by decomposing the velocity $\vec{U}$, pressure $p$, and temperature $T$ fields into the average and fluctuating quantities, $A(\vec{r}, t) = \bar{A}(\vec{r}, t) + a(\vec{r}, t)$, where $\bar{A}$ and $a$ respectively denote the ensemble average $\langle A\rangle$ and fluctuating $a$ terms in the decomposition. Approaches for formulating statistical models of turbulence~\cite{WALLACE2014022003}, such as direct interaction approximation ~\cite{kraichnan_1959, kraichnan_1964} and eddy-damped quasi-normal Markovianization~\cite{lesieur20003d}, proposed evolution equations for higher-order correlations that are obtained by taking moments of the NSE. In these equations, the temporal evolution of correlations, on the left-hand side of the NSE, calls for modeling higher-order moments on the right-hand side of the NSE, where again the strong spatiotemporal correlations between the velocity components $\langle u_i^m u_j^n\rangle$ and between other flow quantities have precluded attempts to close the hierarchy of the moment equations and devise a universal model for turbulent flows.

The availability of massively parallel computing platforms offers a route for direct numerical simulation (DNS) of the Navier--Stokes equations for canonical flows on simple geometries and increasing Reynolds numbers. One such canonical flow that continues to be simulated on newer parallel computing platforms is that of forced isotropic turbulence in a triply periodic box. Using innovative methods to stir the flow (to ensure that the turbulence does not decay), pseudo-spectral solvers for incompressible flows have been used to simulate homogeneous isotropic turbulence (HIT) on boxes consisting of $12288^3$ grid points to achieve a maximum Reynolds number $Re_\lambda = 1300$ (where $\lambda$ denotes the Taylor microscale)\cite{Dhawal2020}. These DNS flow fields can then be used to validate existing theories and  inform the development of subgrid models by exploring the general two-point, two-time, space-time velocity correlation, defined by 
\begin{equation}
    \mathcal{R}(\vec{x}, t) = \langle u_i(\vec{x}, t)u_j(\vec{x}+\vec{r}, t+\tau)\rangle,
\end{equation}
where $u_i, u_j (i, j = 1,2,3)$ denotes the fluctuating velocity components, $\vec{x}+\vec{r} = (x_1+\delta x_1,x_2 + \delta x_2,x_3 +\delta x_3)$ denotes the distance between two points in the flow, $\tau$ denotes the temporal window, and $\langle . \rangle$ denotes the ensemble average. The flow is said to be decorrelated when the correlation decreases with increasing $|\vec{r}|$ and $\tau$ and vanishes above a critical distance $|\vec{r}|_d$ and $\tau_d$ that denote the decorrelation length and time, respectively. Similar expressions can be defined to correlate other quantities, such as the pressure ($p$) and temperature ($T$) between two points and at two different times. 

While  a considerable body of literature exists for HIT simulations of incompressible flows, HIT simulations of compressible flows are not  as extensive. DNS of compressible turbulence requires solving the energy equation in addition to the momentum equations. Another complication that arises in simulating compressible flows is the appearance of shocks and \emph{shocklets} in compressible HIT, where the flow properties change abruptly across the shocklets. Hence, shock-capturing numerical methods are necessary that also resolve all the turbulent scales of motion with minimal dissipation and dispersion errors. Needed are high accuracy, spectral resolution, and the ability to capture unsteady shocklets/shock waves. Weighted essentially nonoscillatory (WENO) schemes~\cite{liuosherchan,jiangshu,shuosher1988,shuosher1989} use solution-dependent reconstruction to achieve high-order accuracy for smooth flows and yield nonoscillatory solution across shocks and other discontinuities. Although WENO schemes of very high orders have been  designed~\cite{balsarashu2000,wuetal2021}, they suffer from relatively poor spectral resolution and are ill-suited for the DNS of turbulent flows. Several strategies were introduced to improve their dissipation and dispersion properties. One strategy is improving the calculation of the nonlinear weights~\cite{wuetal2021,henricketal2005,borgesetal2008,yamaleevcarpenter2009_1,yamaleevcarpenter2009_2} such that they attain their optimal values when the flow is smooth but not well-resolved. Another strategy is improving the spectral resolution of the underlying finite-difference scheme, for example, bandwidth-optimized schemes~\cite{Martin2006}, dissipation-relation-preserving scheme~\cite{sjogreenyee2017}, and minimized dispersion schemes with controllable dissipation~\cite{sunetal2011,sunetal2020,lietal2022,dengetal2022}. Given the high spectral resolution of compact or Pad\'{e}--type schemes~\cite{lele1992}, a related strategy has been using compact finite-difference schemes as the underlying linear discretization; examples include weighted compact schemes (WCNS)~\cite{subramaniametal2019,chenetal2021,hiejima2022} and CRWENO~\cite{ghoshbaeder2012,ghoshbaeder2014,pengshen2015,pengshen2017,fidalgoetal2018} schemes. As an alternative to the WENO schemes, targeted essentially nonoscillatory (TENO) schemes were introduced~\cite{fuetal2016} that modify the stencil selection procedure to reduce the dispersion and dissipation errors at high wavenumbers; they have been successfully applied to the DNS of compressible turbulent flows~\cite{lushersandham2021}. Many efforts combine these ideas in hybrid schemes, for example, hybrid central-upwind finite-difference schemes~\cite{chakravarthyetal2015} and hybrid compact--WENO~\cite{adamshariff1996,pirozzoli2002,renetal2003}.

GPUs require a fundamentally different programming paradigm from that of CPUs. Some memory operations that are efficient on  CPUs could cause a serious performance degradation on GPUs. The past decade has seen several GPU implementations of computational  fluid  dynamics (CFD) codes for simulating turbulent flows. Linear, high-order finite-difference schemes were implemented on GPUs and applied to the DNS of turbulent flows, for example, 4th--order~\cite{jammyetal2019} and 6th--order~\cite{salvadoreetal2013,wangetal2021} central schemes. These studies reported speedups by factors of ~$10$--$20$ when compared with a CPU-only implementation. High-order WENO schemes have been implemented and assessed on GPUs~\cite{esfahanianetal2013,darianesfahanian2014,kuowu2021}, including on one of the fastest supercomputers~\cite{huangetal2019}. One conclusion of these studies was that the WENO kernel was the most expensive component for a typical NSE code. The GPU implementation of WENO-type schemes was applied to the Favre-averaged NSE~\cite{antoniouetal2010,karantasisetal2014}. DNS codes based on WENO-type schemes have been implemented on GPUs to simulate supersonic turbulent flows, such as the shock--turbulence interaction~\cite{postdimare2020}, supersonic channel flow, shock--boundary layer interaction~\cite{bernardinietal2021}, and supersonic compression-expansion slope flow~\cite{xuetal2021}. Given the poor spectral resolution of the traditional WENO scheme, GPU implementations of more modern variants such as the TENO schemes were developed~\cite{hamzehlooetal2021,lusheretal2021,direnzoetal2020}. Given the suitability of compact/pseudo-spectral schemes, a linear 6th-order tridiagonal compact scheme with filtering was implemented for GPUs~\cite{tutkunediz2012} to simulate the evolution of eddies. However, the algorithm suffered from an inefficient implementation of the tridiagonal solve on the GPU that did not scale well with the problem size. Heterogeneous (CPU+GPU) implementations of two shock-capturing compact schemes, the weighted compact nonlinear scheme (WCNS) and hybrid dissipative compact scheme, were developed~\cite{xuetal2014} and applied to simulate flow around aircraft configurations.


The objective of this paper is a computationally efficient heterogeneous (CPU+GPU) implementation of an existing scalable, CPU-parallel shock-capturing CFD algorithm for the DNS of canonical turbulent flows. We report our strategies to accelerate an existing open-source MPI-parallel C/C++ code, \hypar~\cite{hypar}, with GPUs. We consider the fifth-order WENO scheme (WENO5)~\cite{jiangshu}; in subsequent publications, we will consider high-resolution alternatives that are better suited for DNS of turbulent flows, such as the compact-WENO and low-dissipation WENO schemes. Since efficient memory operations are crucial for fully utilizing the massive parallel computing capability of GPUs, our focus is on optimizing the memory access patterns of \hypar{} on GPUs. In particular, we aim to avoid slow memory transactions and reduce as many wasted computational resources (such as thread blocks) and memory transactions as possible. Our optimization approach results in more than 200 times faster computation time using multiple threads on a GPU compared with a CPU core. Using multiple GPUs and CUDA-aware MPI communication, we demonstrate the scalability of our GPU-based implementation. We evaluate strong and weak scaling and investigate the impact of communication cost on the overall computation time. We also simulate the decay of HIT in a periodic domain with up to $1024^3$ grids points ($\sim 5\times10^9$ billion degrees of freedom) on up to $1,024$ GPUs and compare the wall time with MPI-only simulations on up to $8,192$ CPU cores. As a highlight, our current GPU-based implementation can perform 100 time steps on a grid with $1024^3$ ($\sim 10^9$) points in less than 30 seconds using $1,024$ GPUs. 

Our implementation presented here is among the earliest attempts at simulating scale-resolved compressible HIT on DOE leadership-class heterogeneous (\textit{i.e.,} CPU+GPU) platforms. Recent work for incompressible turbulence at such scales includes simulations with up to $\sim21$~billion grid points on $1280$ NVIDIA P100 GPUs~\cite{zolfagharietal2019,zolfaghariobrist2021} and a pseudo-spectral algorithm for grids with up to $18,432^3$ points and $18,432$ NVIDIA V100 GPUs~\cite{ravikumaretal2019}. The simulations presented in this paper were carried out on the Summit supercomputer at the Oak Ridge Leadership Computing Facility (OLCF)~\cite{summit} (referred as ``OLCF/Summit" in the rest of the paper) and the Lassen supercomputer at  Lawrence Livermore National Laboratory~\cite{lassen} (referred as ``LLNL/Lassen"). The implementation described here has been released in the latest version of \hypar~\cite{hypar} under an MIT license.

The rest of the paper is organized as follows. In Section~\ref{sec:background} we introduce our simulation: the governing equations for compressible Navier--Stokes equations, the numerical methods, and the HIT decay example. Section~\ref{sec:impl} describes our kernel design principles for efficiently implementing the existing CPU code-base on GPUs, with a specific focus on optimizing its memory access patterns. In Section~\ref{sec:exp} we report the computational performance and scalability tests of the GPU-based implementation. Section~\ref{sec:results} describes high-resolution simulations of HIT decay on very fine grids and compares the CPU+GPU and CPU-only wall times. We summarize our conclusions in Section~\ref{sec:conclusion}.
\section{Background}
\label{sec:background}

In this section we briefly introduce governing equations for compressible Navier--Stokes equations, their numerical methods, and the isotropic turbulence decay example.
The example will be used to evaluate the computational performance of our GPU-based implementation in Section~\ref{sec:exp}.

\subsection{Governing equations}
The nondimensional, compressible Navier--Stokes equations~\cite{hirsch2007} are expressed as a system of hyperbolic-parabolic partial differential equations  as
\begin{align}\label{eq:NS}
    \frac{\partial {\bf u}}{\partial t}
    +
    \frac{\partial {\bf f}}{\partial x}
    +
    \frac{\partial {\bf g}}{\partial y}
    +
    \frac{\partial {\bf h}}{\partial z}
    =
    \frac{\partial {\bf f}^v}{\partial x}
    +
    \frac{\partial {\bf g}^v}{\partial y}
    +
    \frac{\partial {\bf h}^v}{\partial z},
\end{align}
where ${\bf u} = \left[\rho, \rho u, \rho v, \rho w, e\right]^{\rm T}$ is the vector of conserved variables. The convective fluxes are
\begin{align}
    &{\bf f}
    = \left[\begin{array}{c}
        \rho u\\
        \rho u^2 + p\\
        \rho u v\\
        \rho u w\\
        \left(e+p\right) u
      \end{array}\right],\
    {\bf g}
    = \left[\begin{array}{c}
        \rho v\\
        \rho uv\\
        \rho v^2+p\\
        \rho v w\\
        \left(e+p\right) v
      \end{array}\right],\
    {\bf h}
    = \left[\begin{array}{c}
        \rho w\\
        \rho uw\\
        \rho vw\\
        \rho w^2+p\\
        \left(e+p\right) w
      \end{array}\right],
\end{align}
where $\rho$ is the density, $u,v,w$ are the Cartesian components of the velocity, $p$ is the pressure, and $e$ is the internal energy given by
\begin{align}
    e = \frac{p}{\gamma-1} + \frac{1}{2}\rho\left(u^2+v^2+w^2\right),
\end{align}
and $\gamma = 1.4$ is the heat coefficient ratio. The viscous fluxes are
\begin{align}
    &{\bf f}^v
    = \left[\begin{array}{c}
        0\\
        \tau_{xx}\\
        \tau_{yx}\\
        \tau_{zx}\\
        {\bf v}\cdot\boldsymbol{\tau}_x-q_x
      \end{array}\right],\
    {\bf g}^v
    = \left[\begin{array}{c}
        0\\
        \tau_{xy}\\
        \tau_{yy}\\
        \tau_{zy}\\
        {\bf v}\cdot\boldsymbol{\tau}_y-q_y
      \end{array}\right],\
    {\bf h}^v
    = \left[\begin{array}{c}
        0\\
        \tau_{xz}\\
        \tau_{yz}\\
        \tau_{zz}\\
        {\bf v}\cdot\boldsymbol{\tau}_z-q_z
      \end{array}\right];
\end{align}
the viscous stresses are given by
\begin{align}
    \tau_{ij} = \mu \frac{M_\infty}{Re_\infty} \left[ \left(\frac{\partial u_i}{\partial x_j} - \frac{\partial u_j}{\partial x_i}\right) - \frac{2}{3}\frac{\partial u_k}{\partial x_k}\delta_{ij} \right],
\end{align}
where $Re_\infty$ and $M_\infty$ are the reference Reynolds and Mach numbers, respectively, $\mu$ is the normalized coefficient of viscosity, ${\bf v} \equiv \left(u,v,w\right)$ is the velocity vector, and
\begin{align}
\boldsymbol{\tau}_{\left(\cdot\right)}
=
\left( \tau_{x\cdot}, \tau_{y\cdot}, \tau_{z\cdot} \right).
\end{align}
The thermal conduction terms are
\begin{align}
    q_i = - \frac{\mu}{\left(\gamma-1\right)} \frac{M_\infty}{Re_\infty\ Pr } \frac{\partial T}{\partial x_i},
\end{align}
where $T = \gamma p/\rho$ is the temperature and $Pr=0.72$ is the Prandtl number.

\subsection{Numerical method}

\begin{figure}[t!]
\begin{center}
\includegraphics[width=0.9\textwidth]{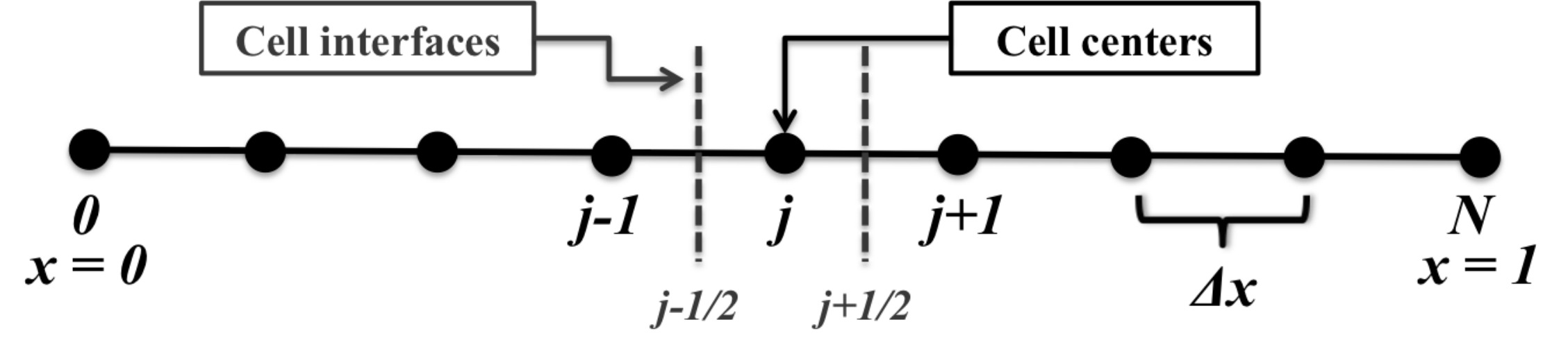}
\caption{Illustration of cell centers and cell interfaces on a grid line along a specific dimension.}
\label{fig:domain1D_disc}
\end{center}
\end{figure}

The physical domain is discretized by  using a three-dimensional grid with $N$ points in each dimension. The convective terms on the left-hand side  of~(\ref{eq:NS}) are discretized with the conservative finite-difference formulation~\cite{shuosher1988,shuosher1989} and the WENO5 scheme~\cite{jiangshu}. This section briefly describes the spatial discretization of the convective flux ${\bf f}$ along $x$; this can be extended trivially to the convective terms along the other dimensions. The spatial derivatives in the viscous terms on the right-hand side (RHS) are discretized by using fourth-order central finite differences. We thus obtain an ordinary differential equation (ODE) in time; this is evolved by using the third-order total-variation-diminishing (TVD) and fourth-order Runge--Kutta methods. A more detailed description of the numerical methodology is available in prior publications~\cite{ghoshbaeder2012,ghoshbaeder2014,ghoshetal2015}.

We describe the upwind discretization of the convective terms by considering an arbitrary grid line along $x$ and ignoring the other dimensions. The derivative is computed at a grid point $j$ as
\begin{align}
    \left.\frac{\partial {\bf f}}{\partial x}\right|_j
    &=
    \frac{1}{\Delta x} \left( \hat{\bf f}_{j+\frac{1}{2}} - \hat{\bf f}_{j-\frac{1}{2}} \right)
    + \mathcal{O}\left(\Delta x^p\right),
\end{align}
where the locations $j\pm 1/2$ denote the cell interfaces (see Figure~\ref{fig:domain1D_disc}) and $p$ is the order of the discretization scheme. The numerical flux at the cell interface, $\hat{\bf f}_{j+\frac{1}{2}}$, is computed by using the Roe scheme~\cite{roe1981} with the Harten entropy fix~\cite{harten1983}:
\begin{align}\label{eq:Roe}
    \hat{\bf f}_{j+\frac{1}{2}}
    =
    \frac{1}{2} \left( \hat{\bf f}_{j+\frac{1}{2}}^L + \hat{\bf f}_{j+\frac{1}{2}}^R \right)
    -
    \frac{1}{2} \left| A_{j+\frac{1}{2}} \right|
    \left( \hat{\bf u}_{j+\frac{1}{2}}^R - \hat{\bf u}_{j+\frac{1}{2}}^L \right),
\end{align}
where the superscripts $L,R$ denote left- and right-biased discretizations, respectively. The dissipation matrix, $\left| A_{j+\frac{1}{2}} \right|$, is computed from the eigensystem evaluated at the Roe-averaged state at the cell interface $j+1/2$ as
\begin{align}
    \left| A_{j+\frac{1}{2}} \right|
    =
    X_{j+\frac{1}{2}}
    \left| \Lambda_{j+\frac{1}{2}} \right|
    X_{j+\frac{1}{2}}^{-1},
\end{align}
where $X$ is the matrix with columns as the right eigenvectors and $\Lambda$ is a diagonal matrix with the eigenvalues as its entries.

Each scalar component of $\hat{\bf f}_{j+\frac{1}{2}}^{L,R}$ and $\hat{\bf u}_{j+\frac{1}{2}}^{L,R}$ is computed by using the WENO5 scheme. We describe this method for a left-biased scalar variable, $\hat{f}_{j+\frac{1}{2}}^L$; the procedure for a right-biased scalar variable, $\hat{\bf f}_{j+\frac{1}{2}}^R$, can be obtained by reflecting the expressions around the cell interface $j+1/2$. We drop the superscript $L$ in the text below.

WENO schemes use a solution-dependent interpolation stencil selection~\cite{liuosherchan,jiangshu} to achieve high-order accuracy where the solution is smooth and to avoid oscillations across discontinuities. The WENO5 scheme is constructed by identifying three third-order interpolation schemes at the cell interface; the final interpolation method is their weighted sum, where the weights depend on the smoothness of the stencils underlying the corresponding third-order interpolation. These weights approach optimal values for smooth solutions; consequently, the method achieves fifth-order  accuracy. In the presence of discontinuities or sharp gradients, the weights corresponding to those stencils approach zero. The final method has a stencil biased away from the discontinuity, thus avoiding numerical oscillations.

The three third-order interpolation schemes and their optimal weights at cell interface $j+1/2$ are
\begin{eqnarray}
\hat{f}_{j+1/2}^1 &=& \frac{1}{3} f_{j-2} - \frac{7}{6} f_{j-1} + \frac{11}{6} f_j,\ c_1 = \frac{1}{10}, \label{eqn:third_order_cand1}\\
\hat{f}_{j+1/2}^2 &=& -\frac{1}{6} f_{j-1} + \frac{5}{6} f_j + \frac{1}{3} f_{j+1},\ c_2 = \frac{6}{10}, \label{eqn:third_order_cand2}\\
\hat{f}_{j+1/2}^3 &=& \frac{1}{3} f_j + \frac{5}{6} f_{j+1} - \frac{1}{6} f_{j+2},\ c_3 = \frac{3}{10}. \label{eqn:third_order_cand3}
\end{eqnarray}
The fifth-order interpolation scheme at $j+1/2$ is obtained as
\begin{align}\label{eqn:fifth_order}
\hat{f}_{j+1/2}
&=
\sum_{k=1,2,3} c_k \hat{f}_{j+1/2}^k\nonumber\\
&=
\frac{1}{30} f_{j-2} - \frac{13}{60} f_{j-1} + \frac{47}{60} f_j + \frac{27}{60} f_{j+1} - \frac{1}{20} f_{j+2}.
\end{align}
The solution-dependent weights are computed as
\begin{equation}\label{eqn:weno_weights}
\omega_k = \frac{\alpha_k}{\sum_k \alpha_k};\ \alpha_k = \frac{c_k}{\left(\epsilon + \beta_k\right)^p};\ k = 1,2,3,
\end{equation}
where $\epsilon=10^{-6}$ is a small number to prevent division by zero and $\beta_k$ are the smoothness indicators for the stencils:
\begin{align}
\beta_1 &= \frac{13}{12} (f_{j-2}-2f_{j-1}+f_{j})^2 + \frac{1}{4}(f_{j-2}-4f_{j-1}+3f_{j})^2, \label{eqn:weno5is1} \\
\beta_2 &= \frac{13}{12} (f_{j-1}-2f_{j}+f_{j+1})^2 + \frac{1}{4}(f_{j-1}-f_{j+1})^2, \label{eqn:weno5is2} \\
\beta_3 &= \frac{13}{12} (f_{j}-2f_{j+1}+f_{j+2})^2 + \frac{1}{4}(3f_{j}-4f_{j+1}+f_{j+2})^2. \label{eqn:weno5is3}
\end{align}
The WENO5 scheme is obtained by multiplying the third-order interpolation schemes by the solution-dependent weights instead of the optimal weights:
\begin{align}\label{eqn:weno5}
\hat{f}_{j+1/2} =
&=
\sum_{k=1,2,3} \omega_k \hat{f}_{j+1/2}^k\nonumber\\
&=
\frac{\omega_1}{3} f_{j-2} - \frac{1}{6}(7\omega_1+\omega_2)f_{j-1} + \frac{1}{6}(11\omega_1+5\omega_2+2\omega_3)f_j\nonumber \\
&\ \ \ \ \ \ \ \ \ \ \ + \frac{1}{6}(2\omega_2+5\omega_3)f_{j+1} - \frac{\omega_3}{6}f_{j+2}.
\end{align}
If the solution is locally smooth, $\omega_k \rightarrow c_k$, $k=1,2,3$, and (\ref{eqn:weno5}) is equivalent to (\ref{eqn:fifth_order}).

\subsection{Example: Isotropic turbulence decay}
\label{subsec:isotropic}

\begin{figure}[t]
    \centering
    \begin{subfigure}[t]{0.49\textwidth}
        \centering
        \includegraphics[width=1.0\textwidth]{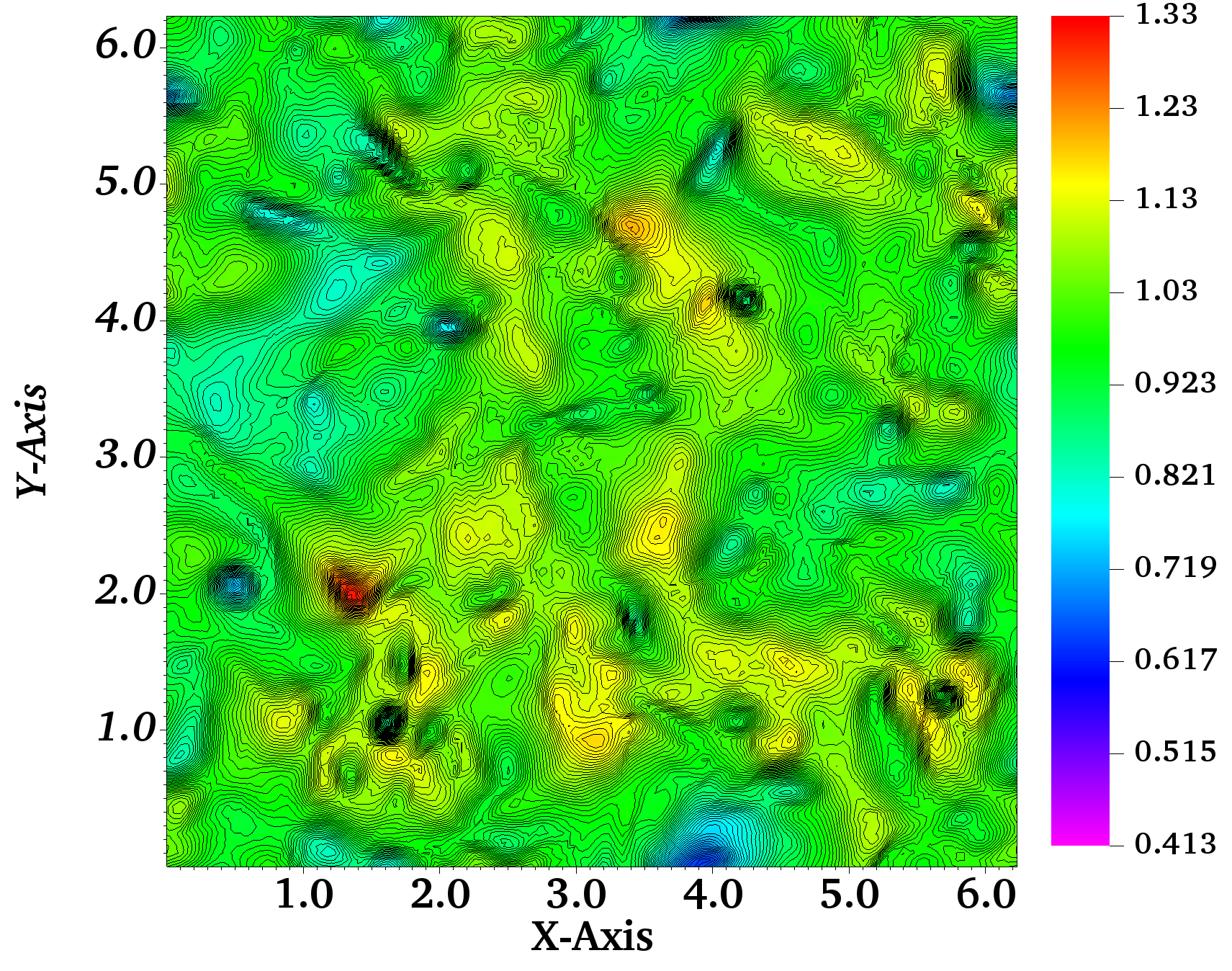}
        \caption{Density at $z=0$ and $t/\tau=3$.}
        \label{fig:density}
    \end{subfigure}
    \begin{subfigure}[t]{0.49\textwidth}
        \centering
        \includegraphics[width=1.0\textwidth]{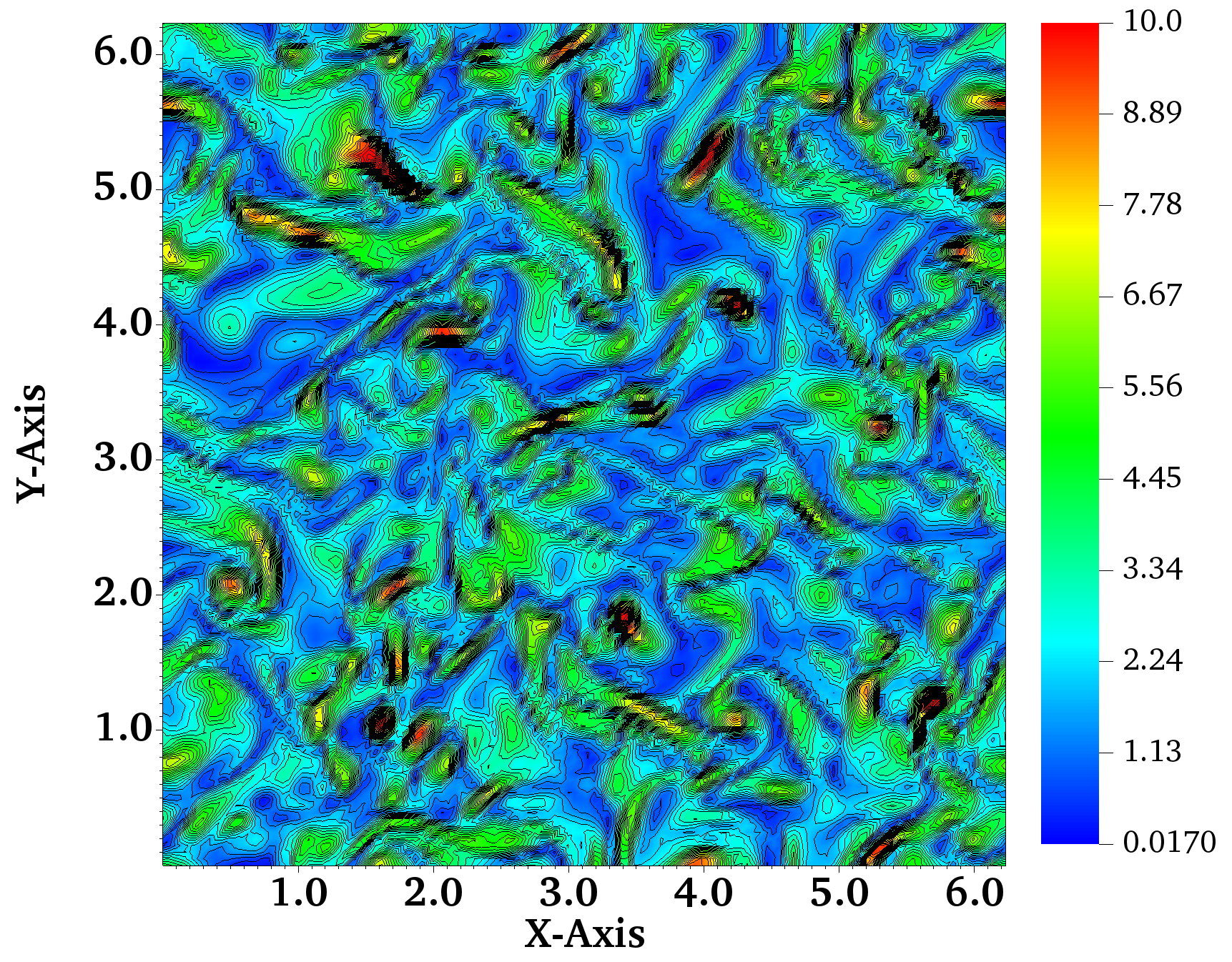}
        \caption{Vorticity magnitude at $z=0$ and $t/\tau=3$.}
        \label{fig:vorticity}
    \end{subfigure}
    \caption{Direct numerical simulation of HIT in a triply periodic domain discretized by a $128^3$ grid.The solution is obtained at a final time of $t/\tau=3$, where $\tau$ is the turbulent time scale.}
\end{figure}

We simulate the decay of isotropic turbulence in a periodic domain; this is a canonical turbulent flow problem~\cite{rogallo1981,mansourwray1994} characterized by the transfer of energy from larger to smaller length scales. Our governing equations do not include a turbulence model; we thus conduct a direct numerical simulation  by resolving the Kolmogorov scale. The initial solution comprises a divergence-free velocity field with random isotropic fluctuations that satisfy the following kinetic energy spectrum:
\begin{align}
    E\left(k\right)
    =
    16 \sqrt{\frac{2}{\pi}} \frac{u_0^2}{k_0}
    \left(\frac{k}{k_0} \right)^4
    \exp\left[ -2 \left(\frac{k}{k_0} \right)^2 \right],
\end{align}
where $k$ is the wavenumber, $E$ is the kinetic energy, $u_0$ is the root-mean-square turbulence intensity, and $k_0$ is the wavenumber corresponding to the maximum energy. The domain is a cube of length $2\pi$ with periodic boundaries. The initial density and pressure are constant ($\rho=1$, $p=1/\gamma$). We specify $u_0=0.3$ and $k_0=4$,  resulting in a smooth turbulent flow. The Taylor microscale-based Reynolds number, $Re_\lambda = \rho u_0 \lambda/\mu$, is specified as $50$, where $\lambda$ is the Taylor microscale.

\begin{figure}[t]
    \centering
    \includegraphics[width=0.9\textwidth]{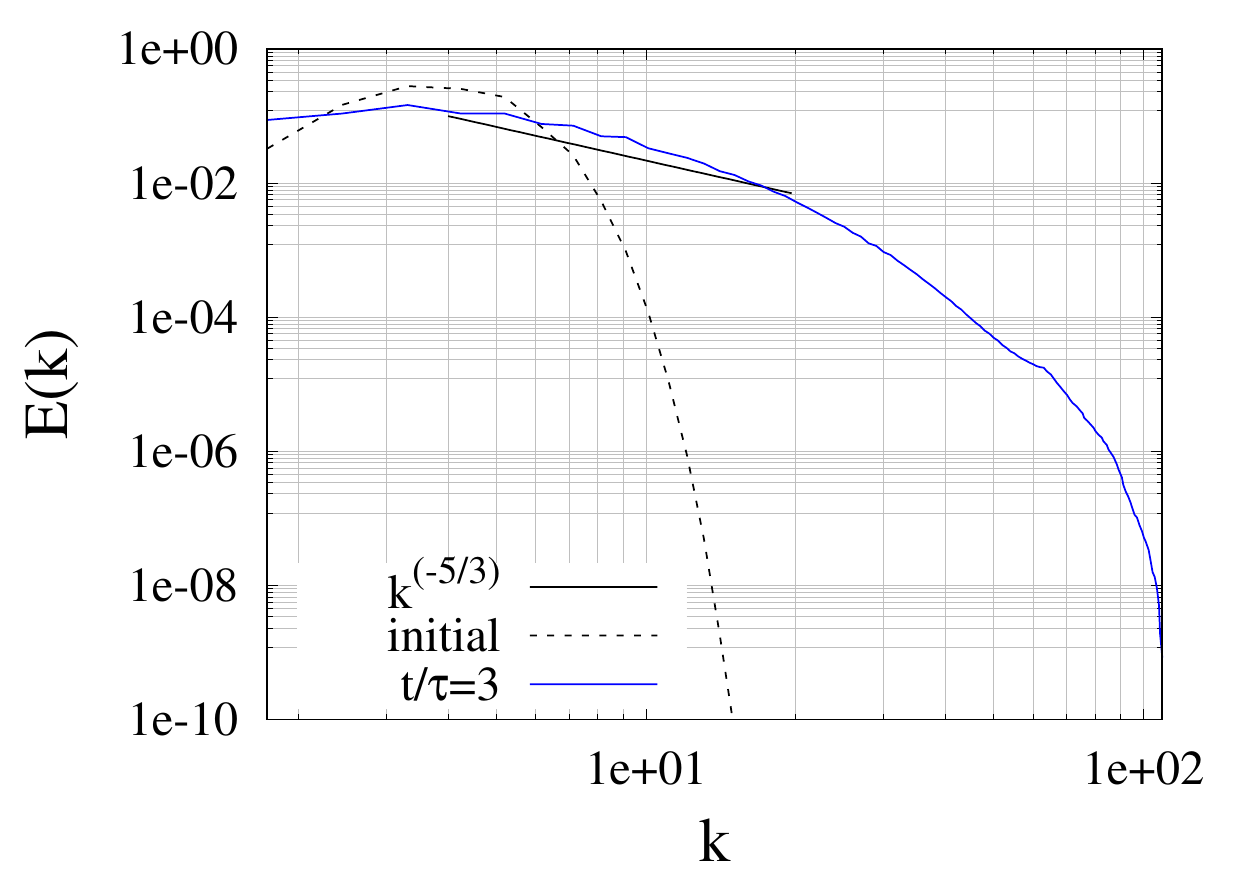}
    \caption{Energy spectrum of the initial and final ($t/\tau=3$) solutions.}
    \label{fig:energy_spectrum}
\end{figure}

Figures~\ref{fig:density} and~\ref{fig:vorticity} show the density and vorticity magnitude contours, respectively, on a two-dimensional plane at $z=0$ for a solution obtained at $t/\tau=3$, where $\tau=\lambda/u_0$ is the turbulent time scale. This solution is obtained on a grid with $128^3$ points and a time step $\Delta t = 0.025$ (CFL $\approx 1$). Figure~\ref{fig:energy_spectrum} shows the initial and final energy spectra for the solution. As the flow evolves and the turbulence decays, we observe an energy transfer from the lower to higher wavenumbers, as expected. The $k^{-5/3}$ relationship is also plotted for reference. We use this example in our subsequent studies.
\section{Implementation Principles on GPUs}
\label{sec:impl}

In this section we present our kernel design principles to efficiently implement \hypar{} on GPUs, with a specific focus on optimizing its memory access patterns.
Since many core operations of \hypar{} are memory-bound, it is crucial for  computational performance to avoid slow memory transactions and to reduce as many wasted compute resources and memory transactions as possible.
To  this end, we introduce the following three design principles that target optimized computational performance on GPUs:
(i) implementing lexicographic thread configuration,
(ii) removing data transfers between CPUs and GPUs during operation,  and
(iii) performing coalesced memory access.
In the following subsections we introduce these principles one by one with numerical performance comparisons with those of alternative approaches.
All of the numerical experiments in this section have been performed on a single NVIDIA's V100 using CUDA.

We note that we have also tried some  user-defined caching approaches, such as manually allocating dynamic shared memory (instead of relying on the underlying GPU's automatic caching mechanism) to share function values in~\eqref{eqn:weno5is1}--\eqref{eqn:weno5is3} between different grid points.
However, we observed no gains or even worse performance than when we relied on the given automatic caching mechanism.
Therefore, we did not employ manual shared-memory management in our implementation.
We believe that such a worse performance with shared memory is partly due to a significantly improved cache performance of the Volta architecture, as described in~\cite[See page 17]{Volta}.

\subsection{Lexicographic thread block configuration}
\label{subsec:thread-config}

By default, we have a one-to-one mapping from threads of a GPU kernel into the grid points of a discretization scheme so that each thread is assigned to a single grid point.
One way to implement this mapping is to assign threads in the same lexicographic order of the coordinate values of the grid points.
In this case, each thread block is one-dimensional, and the order of threads within a block follows the same lexicographic order of the corresponding grid points.
An alternative way to implement the mapping is to use tiling.
For example, if a grid is $64 \times 64$, we can generate thread blocks, each of which is of size $32 \times 8$ (256 threads per block).
A total of 16 thread blocks are needed to cover the entire grid in this case.

The tiling approach is intuitive; however, it could cause many wasted threads when some dimension of the grid is not a multiple of the number of threads of a warp.
For example, if a given grid is of size $65 \times 64$, the previous $32 \times 8$ thread block scheme will result in some thread blocks having many inactive threads.
In this case, thread blocks covering grid points with 65 as its $x$-coordinate have only 8 active threads among 256 threads.
Such an unfavorable grid naturally occurs when we perform computation using ghost points, which are added to the existing grid scheme.

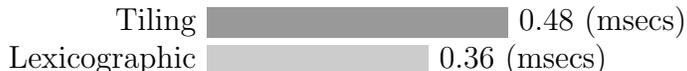
\begin{figure}[htbp]
    \centering
    \begin{tikzpicture}
    \node[left] at (0,1.5) {Tiling};
    \node[left] at (0,1.0) {Lexicographic};
    
    \fill[gray!80] (0,1.3) rectangle (4,1.7);
    \node[right] at (4,1.5) {0.48 (msecs)};
    
    \fill[gray!40] (0,0.8) rectangle (4*.7359,1.2);
    \node[right] at (4*.7359,1.0) {0.36 (msecs)};
    \end{tikzpicture}
    \caption{Tiling vs lexicographic over weights computation~\eqref{eqn:weno_weights}.}
    \label{fig:tiling-lex}
\end{figure}

Figure~\ref{fig:tiling-lex} compares the computation time between the tiling and lexicographic thread block configurations, obtained from a kernel computing the weights~\eqref{eqn:weno_weights}.
The grid employed is of size $65 \times 64 \times 64$, and the thread block configuration for the tiling approach is $32 \times 8$.
As we see in the figure, the tiling approach shows a slower computation time than that of the lexicographic configuration.
Similar results were obtained with different block configurations for tiling.

\subsection{No data transfers between CPUs and GPUs}
\label{subsec:no-data-transfer}

\begin{figure}[htbp]
    \centering
    \begin{tikzpicture}
    \foreach \l/\x/\p[count=\y] in {{\tt Derivative}/.3072/5.12, {\tt Parabolic}/.5718/9.53, {\tt Upwind}/.69/11.50, {\tt Interp}/1.1136/18.56, {\tt WENO}/2.7012/45.02} {
        \node[left] at (0,.5*\y) {\l};
        \fill[gray!80] (0,.5*\y-.2) rectangle (\x,.5*\y+.2);
        \node[right] at (\x,.5*\y) {\p\%};
    }
    \foreach \x/\p in {1.2/20, 2.4/40, 3.59/60, 4.8/80, 6/100}  {
        \draw (\x,.2) -- (\x,0) node[below] {\p\%};
    }
    \draw (0,0) -- (6,0);
    \draw (0,0) -- (0,2.8);
    \end{tikzpicture}
    \caption{Percentage of time spent on core operations.}
    \label{fig:profiling}
\end{figure}
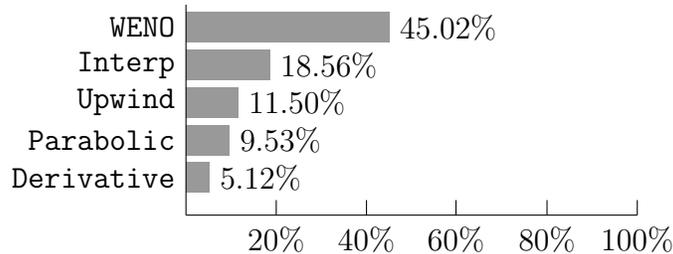

When we solve the three-dimensional Navier--Stokes equations~\eqref{eq:NS} using the WENO5 scheme, our profiling results\footnote{We used {\tt gprof} available at \url{https://sourceware.org/binutils/docs/gprof}.} described in Figure~\ref{fig:profiling} show that about 90\% of the computation time has been spent in the five core operations:  {\tt WENO} -- computing weights~\eqref{eqn:weno_weights}, {\tt Interp} -- calculation of the flux at cell interfaces~\eqref{eqn:weno5}, {\tt Upwind} -- computing the upwind flux using Roe's scheme~\eqref{eq:Roe}, {\tt Parabolic} -- computing the viscous terms on the RHS of~\eqref{eq:NS}, and {\tt Derivative} -- computing the derivatives in the viscous terms using fourth-order central finite differences.
Since these operations are defined for each grid point independently, they can be computed in parallel, providing an opportunity for acceleration on GPUs.
Our initial implementation of computing weights ({\tt WENO}) on GPUs shows about two orders of magnitude of faster computation time than on CPUs.

\begin{figure}[htbp]
    \centering
    \begin{tikzpicture}
    \node[left] at (0,1.5) {CPUs};
    \node[left] at (0,1.0) {GPUs};
    \node[left] at (0,.5)  {Data transfer};
    
    \fill[gray!80] (0,1.3) rectangle (4*.5654,1.7);
    \node[right] at (4*.5654,1.5) {157.43 (msecs)};
    
    \fill[gray!40] (0,0.8) rectangle (4*.0012,1.2);
    \node[right] at (4*.0012,1.0) {0.36 (msecs)};
    
    \fill[gray!40] (0,.3) rectangle (4,.7);
    \node[right] at (4,.5) {272.27 (msecs)};
    \end{tikzpicture}
    \caption{Computation and data transfer times for {\tt WENO}.}
    \label{fig:data-transfer-time}
\end{figure}
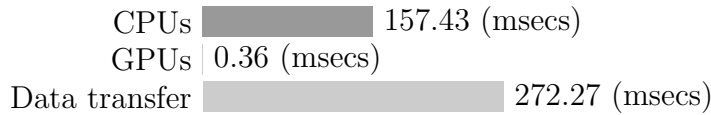

However, if we include data transfer time to initiate the computation on the device and return the results back to the host, the overall computation time significantly degrades.
In Figure~\ref{fig:data-transfer-time} we present the computation times on CPUs and GPUs and data transfer time for one function call of the {\tt WENO} routine for $64 \times 64 \times 64$ grid.
Although the computation time has improved by more than 400 times on GPUs, the time for data transfer dominates the overall computation time, making it even slower than on CPUs.
The high data transfer time is attributed to the large memory size for storing weights, which is around 400 MB.

To avoid such significant degradation, we implement all operations that process data entirely on GPUs, except for the initial data read and return of a final solution.
Since data required for each GPU kernel is processed only on GPUs, they can be readily found in the GPU's global memory.
Therefore, we do not have to transfer any data between CPUs and GPUs during operation, removing one of the main performance bottlenecks.
This allows us to recover the fast computation time on GPUs as shown in Figure~\ref{fig:data-transfer-time}.
We note that we also used CUDA-aware MPI for MPI implementations to support multiple GPUs so that we do not have to communicate via staging through CPUs.

\subsection{Coalesced memory access}
\label{subsec:coalesced-access}

In Figure~\ref{fig:old-memory-layout} we show the memory layout of our original CPU-based code, where we use double precision for floating-point computation.
In the figure, each coordinate value of $(\bar{x},\bar{y},\bar{z})$ denotes the maximum grid coordinate value of the corresponding dimension, and the superscript/subscript on $v$ represents the index of a grid point/variable, respectively.
The five variables for each grid point are located one after the other in memory.
Because grid points are allocated in memory in lexicographic order and are accessed sequentially along with their five variables, memory accesses exhibit locality in this case, leading to fast memory operations through a high cache hit ratio.

\begin{figure}[htbp]
    \centering
    \begin{subfigure}{1\columnwidth}
    \centering
        \begin{tikzpicture}
        \draw (0,0) rectangle (10.2,.5);
        \node at (0,.7) {0};
        \foreach \i[evaluate=\i as \x using \i*.8] in {1,2,...,5} {
            \draw (\x,.5) -- (\x,0);
            \node at (\x-.5,.25) {$v^1_{\i}$};
        }
        \draw [decorate,decoration={brace,mirror,raise=.5ex}] (0,0) -- (4,0) node[midway,yshift=-1.2em] {(0,0,0)};
        \foreach \i[evaluate=\i as \x using (\i/8)*.8] in {8,16,...,40} {
            \node at (\x,.7) {\i};
        }
        \node at (5,.25) {...};
        \foreach \i[evaluate=\i as \x using 6.0+(\i*.8)] in {1,2,...,5} {
            \draw (\x-.5,.5) -- (\x-.5,0);
            \node at (\x-.3,.25) {$v^n_{\i}$};
        }
        \draw [decorate,decoration={brace,mirror,raise=.5ex}] (6.3,0) -- (10.2,0) node[midway,yshift=-1.2em] {($\bar{x},\bar{y},\bar{z}$)};
        \end{tikzpicture}
    \caption{Old memory layout (CPU-friendly).}
    \label{fig:old-memory-layout}
    \end{subfigure}
    \begin{subfigure}{1\columnwidth}
    \centering
        \begin{tikzpicture}
        \draw (0,0) rectangle (10.3,.5);
        \node at (0,.7) {0};
        \foreach \i[evaluate=\i as \x using \i*.8] in {1,2,...,5} {
            \draw (\x,.5) -- (\x,0);
            \node at (\x-.5,.25) {$v^{\i}_1$};
        }
        \foreach \i[evaluate=\i as \x using (\i/8)*.8] in {8,16,...,40} {
            \node at (\x,.7) {\i};
        }
        \node at (5,.25) {...};
        \foreach \i[evaluate=\i as \x using 6.0+(\i*.8)] in {1,2,...,5} {
            \draw (\x-.5,.5) -- (\x-.5,0);
        }
        \foreach \i[evaluate=\i as \x using 9.9-(\i*.8)] in {4,3,2,1} {
            \node at (\x,.25) {$v^{n-\i}_5$};
        }
        \node at (9.9,.25) {$v^n_5$};
        \end{tikzpicture}
    \caption{New memory layout (GPU-friendly).}
    \label{fig:new-memory-layout}
    \end{subfigure}
    \caption{Memory layout: old vs new.}
    \label{fig:memory-layout}
\end{figure}

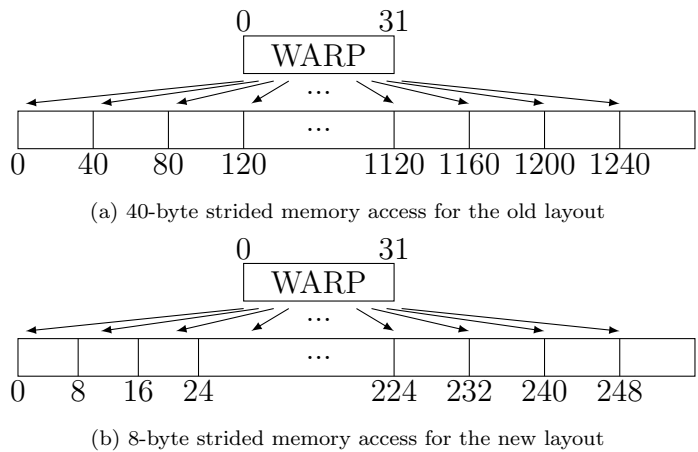
\begin{figure}[htbp]
    \centering
    \begin{subfigure}{1\columnwidth}
    \centering
        \begin{tikzpicture}
        \node at (3,1.7) {0};
        \node at (5,1.7) {31};
        \draw (3,1) rectangle (5,1.5);
        \node at (4,1.25) {WARP};
        \foreach \i[evaluate=\i as \w using 3+\i*.2, evaluate=\i as \x using .1+\i*1.0] in {0,1,...,3} {
            \draw[-latex] (\w,.9) -- (\x,.6);
        }
        \foreach \i[evaluate=\i as \w using 4.5+\i*.2, evaluate=\i as \x using 5.0+\i*1.0] in {0,1,...,3} {
            \draw[-latex] (\w,.9) -- (\x,.6);
            \draw (\x,.5) -- (\x,0);
        }
        \node at (4,.75) {...};
        \node at (4,.25) {...};
        \node at (5.0,-.2) {1120};
        \node at (6.0,-.2) {1160};
        \node at (7.0,-.2) {1200};
        \node at (8.0,-.2) {1240};
        \draw (0,0) rectangle (9,.5);
        \node at (0,.-.2) {0};
        \foreach \i[evaluate=\i as \x using (\i/40)*1.0] in {40,80,120} {
            \draw (\x,.5) -- (\x,0);
            \node at (\x,-.2) {\i};
        }
        \end{tikzpicture}
        \caption{40-byte strided memory access for the old layout}
        \label{fig:old-strided-access}
    \end{subfigure}
    \begin{subfigure}{1\columnwidth}
    \centering
        \begin{tikzpicture}
        \node at (3,1.7) {0};
        \node at (5,1.7) {31};
        \draw (3,1) rectangle (5,1.5);
        \node at (4,1.25) {WARP};
        \foreach \i[evaluate=\i as \w using 3+\i*.2, evaluate=\i as \x using .1+\i*1.0] in {0,1,...,3} {
            \draw[-latex] (\w,.9) -- (\x,.6);
        }
        \foreach \i[evaluate=\i as \w using 4.5+\i*.2, evaluate=\i as \x using 5.0+\i*1.0] in {0,1,...,3} {
            \draw[-latex] (\w,.9) -- (\x,.6);
            \draw (\x,.5) -- (\x,0);
        }
        \node at (4,.75) {...};
        \node at (4,.25) {...};
        \node at (5.0,-.2) {224};
        \node at (6.0,-.2) {232};
        \node at (7.0,-.2) {240};
        \node at (8.0,-.2) {248};
        \draw (0,0) rectangle (9,.5);
        \node at (0,.-.2) {0};
        \foreach \i[evaluate=\i as \x using (\i/8)*.8] in {8,16,24} {
            \draw (\x,.5) -- (\x,0);
            \node at (\x,-.2) {\i};
        }
        \end{tikzpicture}
        \caption{8-byte strided memory access for the new layout}
        \label{fig:new-strided-access}
    \end{subfigure}
    \caption{Different strides for different memory layouts.}
    \label{fig:strided-access}
\end{figure}

However, this memory layout gives rise to 40-byte strided memory accesses on GPUs, as shown in Figure~\ref{fig:old-strided-access}, potentially incurring more memory transactions than needed.
Unlike the CPU case, 32 threads in a warp are executed in lockstep on GPUs.
Since each grid point has five variables of size 8 bytes each (double precision), a warp accessing the first variable of 32 grid points will result in 10 128-byte memory transactions\,footnote{The granularity of L1 cache update is 128-byte.} assuming that the accessed data have not been found in cache.
Bus utilization (the ratio of data actually read to fetched) is 20\% $(=256/1280\times 100)$ in this case.

In contrast, if we allocate each variable of grid points one after the other as shown in Figure~\ref{fig:new-memory-layout}, we will need only 2 128-byte memory transactions with 100\% bus utilization $(=256/256 \times 100)$, as shown in Figure~\ref{fig:new-strided-access}.
Therefore, coalesced memory access provides more efficient memory transactions.

Although bus utilization is low in the former case, later accesses to other variables might be served via cache, resulting in similar overall memory access time to that of the latter case.
However, this expectation depends on the lifetime of fetched data in cache.
The amount of fetched data is proportional to the number of grid points.
Since cache size is limited (128 KB for L1 cache on NVIDIA's Volta architecture), for larger numbers of grid points the likelihood of the eviction of data in cache becomes higher, reducing their lifetime in cache.

\begin{figure}[htbp]
    \centering
    \begin{tikzpicture}
    \begin{axis}[
        scale=.9,
        xlabel={Grid size},
        ylabel={Memory bandwidth (GB/s)},
        xlabel near ticks,
        ylabel near ticks,
        ymin=0, ymax=500,
        xmode=normal,
        xmax=4.4,
        xtick={1,2,3,4},
        xticklabels={$16^3$,$32^3$,$48^3$,$64^3$},
        xmajorgrids=true,
        ymajorgrids=true,
        grid style={line width=.1pt, draw=gray!30},
        legend style={legend pos=north west,cells={anchor=east}},
    ]
        
    \addplot[
        color=black,
        mark=square,
        mark options={scale=2,solid},
    ] coordinates {
        (1,46.21)(2,227.11)(3,326.67)(4,388.22)
    };
    \addlegendentry{8-byte strided access};
    
    \addplot[
        color=black,
        mark=diamond,
        mark options={scale=2,solid},
    ] coordinates {
        (1,44.18)(2,99.70)(3,100.17)(4,106.34)
    };
    \addlegendentry{40-byte strided access};
    
    \draw[dashed,-latex] (axis cs:2,99.70)  -- node[right]{2.2x} (axis cs:2,227.11);
    \draw[dashed,-latex] (axis cs:3,100.17) -- node[right]{3.2x} (axis cs:3,326.67);
    \draw[dashed,-latex] (axis cs:4,106.34) -- node[right]{3.6x} (axis cs:4,388.22);
    \end{axis}
    \end{tikzpicture}
    \caption{Effect of different strides on memory bandwidth.}
    \label{fig:effect-strided-access}
\end{figure}
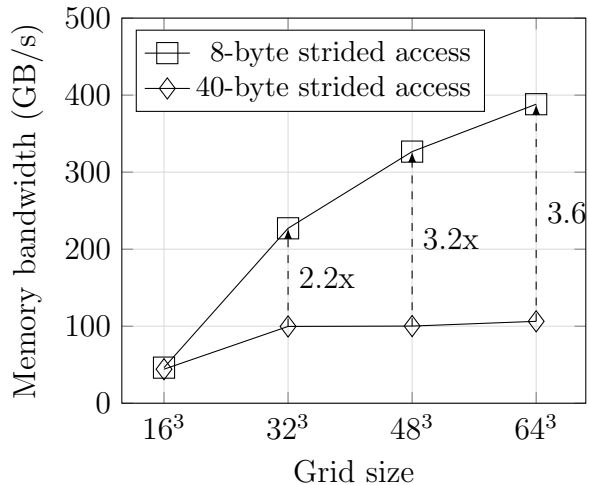

Figure~\ref{fig:effect-strided-access} demonstrates the effect of the aforementioned two strided accesses on memory bandwidth with respect to varying grid sizes.
In the figure, the $x$-label represents the size of each dimension; for example, 16 represents $16=|x|=|y|=|z|$.
Since we have the same number of grid points along each dimension, the total number of grid points is $16 \times 16 \times 16$ in this case.
The memory bandwidth was measured on the kernel to compute the weights~\eqref{eqn:weno_weights}.
We observe that the speedup of 8-byte strided access with respect to 40-byte strided access becomes larger as we increase the grid size.
If data were frequently found in cache for 40-byte strided access, the bandwidth ratio between those two access methods would be similar and stay constant.
However, the speedup is amplified as we increase the grid size, suggesting that there have been many more wasted memory transactions for 40-byte strided access than those of 8-byte strided access.

\section{Computational Performance}
\label{sec:exp}

In this section we present the computational performance of our GPU-based implementation on the isotropic turbulence decay example described in Section~\ref{subsec:isotropic}.
In Section~\ref{subsec:exp-one-gpu} we compare its computation time with that of the CPU-based implementation.
We demonstrate the scalability of our implementation using multiple GPUs in Section~\ref{subsec:exp-multiple-gpus}.

\subsection{Setting}
\label{subsec:exp-setting}

We use the same parameters for the isotropic turbulence decay example as described in Section~\ref{subsec:isotropic} except for time step, which we set to 0.002.
Since we are interested mainly in the speedup of the computation time of our GPU-based implementation, we perform 100 time steps and measure the wall time for all experiments in this section.

The experiments in this section were performed on OLCF/Summit; each compute node has 6 NVIDIA Volta V100 GPUs and 2 POWER9 CPU sockets with 22 physical cores each and 4 hardware threads on each physical core.

\subsection{One GPU vs. one CPU core}
\label{subsec:exp-one-gpu}

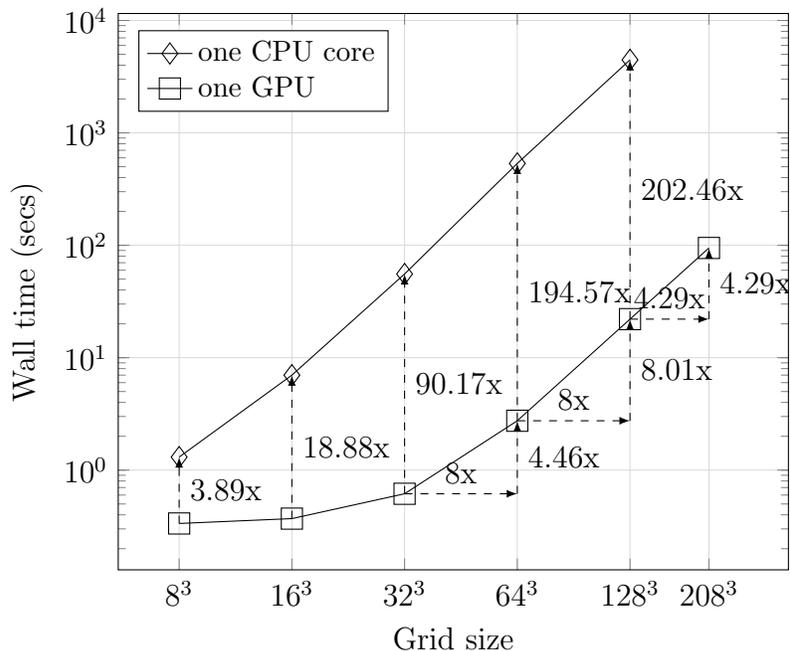
\begin{figure}[t]
    \centering
    \begin{tikzpicture}
    \begin{loglogaxis}[
        scale=1.3,
        xlabel={Grid size},
        ylabel={Wall time (secs)},
        xmax=38998912,
        xtick={512,4096,32768,262144,2097152,8998912},
        xticklabels={$8^3$,$16^3$,$32^3$,$64^3$,$128^3$,$208^3$},
        xmajorgrids=true,
        ymajorgrids=true,
        xlabel near ticks,
        ylabel near ticks,
        grid style={line width=.1pt, draw=gray!30},
        legend style={legend pos=north west,cells={anchor=west}},
    ]

    \addplot [
        color=black,
        mark=diamond,
        mark options={scale=2,solid},
    ] coordinates {
        (512,1.3064)(4096,6.9964)(32768,55.5590)(262144,534.8109)(2097152,4461.0728)
    };
    \addlegendentry{one CPU core}

    \addplot [
        color=black,
        mark=square,
        mark options={scale=2,solid},
    ] coordinates {
        (512,.3353)(4096,.3704)(32768,.6161)(262144,2.7486)(2097152,22.0333)(8998912,94.6541)
    };
    \addlegendentry{one GPU}
    
    \draw[dashed,-latex] (axis cs:512,.3353) -- node[right]{3.89x} (axis cs:512,1.3064);
    \draw[dashed,-latex] (axis cs:4096,.3704) -- node[right]{18.88x} (axis cs:4096,6.9964);
    \draw[dashed,-latex] (axis cs:32768,.6161) -- node[right]{90.17x} (axis cs:32768,55.5590);
    \draw[dashed,-latex] (axis cs:262144,2.7486) -- node[right]{194.57x} (axis cs:262144,534.8109);
    \draw[dashed,-latex] (axis cs:2097152,22.0333) -- node[right]{202.46x} (axis cs:2097152,4461.0728);
    
    \draw[dashed,-latex] (axis cs:32768,.6161) -- node[above]{8x} (axis cs:262144,.6161);
    \draw[dashed,-latex] (axis cs:262144,.6161) -- node[right]{4.46x} (axis cs:262144,2.7486);
    \draw[dashed,-latex] (axis cs:262144,2.7486) -- node[above]{8x} (axis cs:2097152,2.7486);
    \draw[dashed,-latex] (axis cs:2097152,2.7486) -- node[right]{8.01x} (axis cs:2097152,22.0333);
    \draw[dashed,-latex] (axis cs:2097152,22.0333) -- node[above]{4.29x} (axis cs:8998912,22.0333);
    \draw[dashed,-latex] (axis cs:8998912,22.0333) -- node[right]{4.29x} (axis cs:8998912,94.6541);
    
    \end{loglogaxis}
    \end{tikzpicture}    
    \caption{One GPU vs one CPU core for 100 time steps.}
    \label{fig:exp-cpu-vs-gpu}
\end{figure}

Figure~\ref{fig:exp-cpu-vs-gpu} presents the wall times of one GPU and one CPU core, respectively, for executing 100 time steps over varying grid sizes.
From the results, we make two observations.
The first observation is that up until $64^3$ the grid size is not large enough to fully utilize the massive parallel computing capability of GPUs.
Although the grid size becomes 8 times larger when we change its size from $32^3$ to $64^3$, the computation time on GPUs increases by just 4.46x.
This behavior is in contrast to the proportional increase of wall times on CPUs as we increase the grid size.
The disproportional increase of the computation time on GPUs between grids with $32^3$ and $64^3$ points implies that there are many  idle computation units when we solve over grids with $32^3$ points or less.

Once we reach a grid with $64^3$ points, however, the computational resources on GPUs start to saturate.
This is verified by checking the computation time between grids of sizes $64^3$ and $128^3$: the wall time increases exactly at the same rate as the increase of the grid size in this case.
As shown in the figure, for grids having more than $64^3$ points this proportionality is maintained, implying that we are extensively making use of GPUs.
In the case where multiple GPUs are employed as in Section~\ref{subsec:exp-multiple-gpus}, our observation provides a guideline for determining the appropriate number of grid points to be assigned to each GPU that prevents underutilization.

We note that a grid with $208^3$ points is the maximum size that \hypar{} can run on one GPU because of the 32 GB memory limit on V100.
For V100 with 16 GB of memory, the maximum size is around $160^3$.
We also note that we did not measure the computation time on the CPU for the largest grid, since it is expected to be more than 5 hours based on its consistent increase in computation time as the grid size increases.

Our second observation is that the computation time with multiple threads on one GPU is about 200 times faster than that on one CPU core.
The speedup is largely attributed to the acceleration of the {\tt WENO} routine: we achieved about 400 times faster computation time compared with that of the CPU-based code.
Since it takes about 45\% of the total computation time on CPUs as shown in Figure~\ref{fig:profiling}, the improved computation time of the {\tt WENO} routine significantly affects the overall acceleration on GPU.
However, not all routines showed such large gains.
For the {\tt Interp} and {\tt Upwind} routines, the gains were about a factor of 100 and 250, respectively.
The reason is that they are computationally less involving than the {\tt WENO} routine.

We note that the accelerations for smaller grids are not as large as those for larger grids as shown in Figure~\ref{fig:exp-cpu-vs-gpu}, since they are not large enough to fully utilize GPUs, as we have seen in the preceding observation.

\subsection{Multiple GPUs using MPI}
\label{subsec:exp-multiple-gpus}
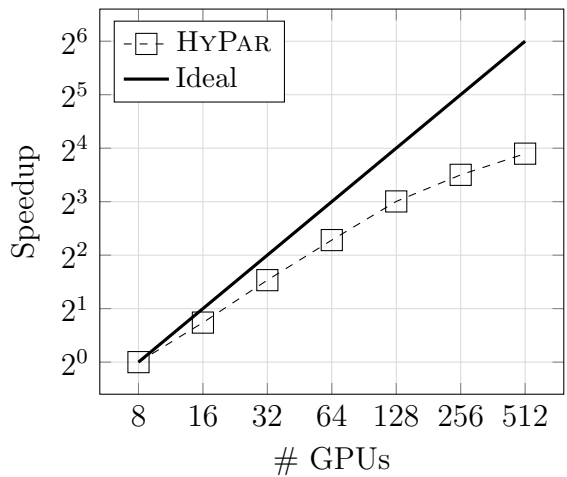
\begin{figure}[t]
    \centering
    \begin{tikzpicture}
    \begin{loglogaxis}[
        scale=.9,
        xlabel={\# GPUs},
        ylabel={Speedup},
        xtick={8,16,32,64,128,256,512},
        xticklabels={8,16,32,64,128,256,512},
        xmajorgrids=true,
        ymajorgrids=true,
        ylabel near ticks,
        xlabel near ticks,
        log base y={2},
        grid style={line width=.1pt, draw=gray!30},
        legend style={legend pos=north west,cells={anchor=west}},
    ]
    
    \addplot[
        dashed,
        color=black,
        mark=square,
        mark options={scale=2,solid},
    ] coordinates {
        (8,24.547897/24.547897)(16,24.547897/14.706084)(32,24.547897/8.485621)(64,24.547897/5.048896)(128,24.547897/3.059928)(256,24.547897/2.167616)(512,24.547897/1.647811)
    };
    \addlegendentry{\hypar{}}
    
    \addplot[
        color=black,
        very thick
    ] coordinates {
        (8,1)(16,2)(32,4)(64,8)(128,16)(256,32)(512,64)
    };
    \addlegendentry{Ideal}
    \end{loglogaxis}
    \end{tikzpicture}
    \caption{Strong scaling over a grid with $256^3$ points.}
    \label{fig:exp-strong-scaling}
\end{figure}

We demonstrate scalability of \hypar{} by evaluating strong and weak scaling over varying numbers of GPUs and grid points.
Communication between GPUs was performed via CUDA-aware MPI, which allowed us to directly communicate between memories of multiple GPUs on the same or different nodes without staging them through CPUs.

\begin{table}[t]
\centering
    \caption{Grid shape per GPU for decomposing $256^3$ grid}
    \label{tbl:exp-grid-shape}
    \begin{tabular}{r|r}
    \hline
        \multicolumn{1}{c|}{\# GPUs} & \multicolumn{1}{c}{Grid shape per GPU}\\\hline
         8 & $128 \times 128 \times 128$ \\
        16 & $64 \times 128 \times 128$ \\
        32 & $64 \times 64 \times 128$ \\
        64 & $64 \times 64 \times 64$ \\
        128 & $32 \times 64 \times 64$ \\
        256 & $32 \times 32\times 64$ \\
        512 & $32 \times 32 \times 32$ \\
        \hline
    \end{tabular}
\end{table}

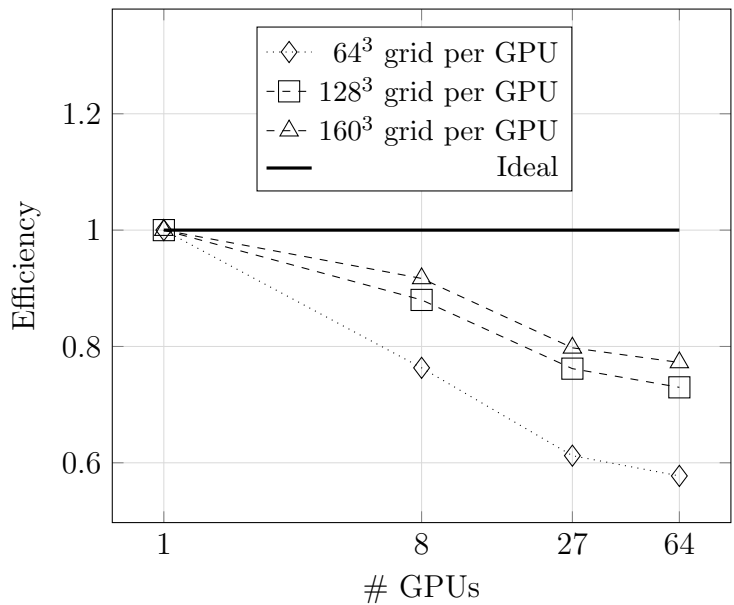
\begin{figure}[t]
    \centering
    \begin{tikzpicture}
    \begin{semilogxaxis}[
        scale=1.2,
        xlabel={\# GPUs},
        ylabel={Efficiency},
        ymax=1.38,
        xtick={1,8,27,64},
        xticklabels={1,8,27,64},
        xmajorgrids=true,
        ymajorgrids=true,
        log base x={2},
        ylabel near ticks,
        xlabel near ticks,
        grid style={line width=.1pt, draw=gray!30},
        legend style={at={(axis cs:27,1.35)},cells={anchor=east}},
    ]

    \addplot [
        dotted,
        color=black,
        mark=diamond,
        mark options={scale=2,solid},
    ] coordinates {
        (1,2.915458/2.915458)(8,2.915458/3.819799)(27,2.915458/4.762164)(64,2.915458/5.048896)
    };
    \addlegendentry{$64^3$ grid per GPU}
    
    \addplot[
        dashed,
        color=black,
        mark=square,
        mark options={scale=2,solid},
    ] coordinates {
        (1,21.601510/21.601510)(8,21.601510/24.547897)(27,21.601510/28.354450)(64,21.601510/29.602674)
    };
    \addlegendentry{$128^3$ grid per GPU}
    
    \addplot[
        dashed,
        color=black,
        mark=triangle,
        mark options={scale=2,solid},
    ] coordinates {
        (1,42.423836/42.423836)(8,42.423836/46.266260)(27,42.423836/53.186267)(64,42.423836/54.918442)
    };
    \addlegendentry{$160^3$ grid per GPU}
    
    \addplot[
        color=black,
        very thick
    ] coordinates {
        (1,1)(8,1)(27,1)(64,1)
    };
    \addlegendentry{Ideal}
    
    \end{semilogxaxis}
    \end{tikzpicture}
    \caption{Weak scaling over up to 64 GPUs and $640^3$ grid}
    \label{fig:exp-weak-scaling}
\end{figure}

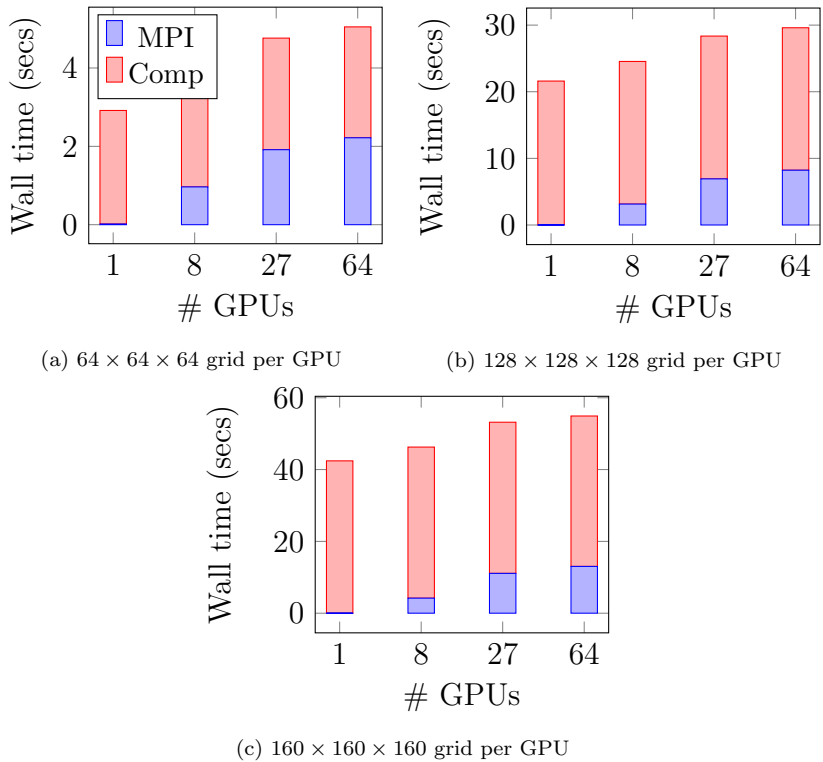
\begin{figure}[t]
    \centering
    \begin{subfigure}{.4\textwidth}
        \centering
        \begin{tikzpicture}
        \begin{axis}[
            width={\textwidth},
            ybar stacked,
            xlabel={\# GPUs},
            ylabel={Wall time (secs)},
            ylabel near ticks,
            xtick={1,2,3,4},
            xticklabels={1,8,27,64},
            legend style={legend pos={north west}},
        ]

        \addplot coordinates {
            (1,0.017699)(2,0.964652)(3,1.914213)(4,2.218241)
        };
        \addlegendentry{MPI}
        
        \addplot coordinates {
            (1,2.915458-0.017699)(2,3.819799-0.964652)(3,4.762164-1.914213)(4,5.048896-2.218241)
        };
        \addlegendentry{Comp}
        
        \end{axis}
        \end{tikzpicture}
    \caption{$64 \times 64 \times 64$ grid per GPU}    
    \end{subfigure}
    \begin{subfigure}{.4\textwidth}
        \centering
        \begin{tikzpicture}
        \begin{axis}[
            width={\textwidth},
            ybar stacked,
            xlabel={\# GPUs},
            ylabel={Wall time (secs)},
            xtick={1,2,3,4},
            xticklabels={1,8,27,64},
        ]
        
        \addplot coordinates {
            (1,0.017731)(2,3.141714)(3,6.922252)(4,8.223817)
        };
        
        \addplot coordinates {
            (1,21.601510-0.017731)(2,24.547897-3.141714)(3,28.354450-6.922252)(4,29.602674-8.223817)
        };
        
        \end{axis}
        \end{tikzpicture}
    \caption{$128 \times 128 \times 128$ grid per GPU}
    \end{subfigure}
    
    \begin{subfigure}{.4\textwidth}
        \centering
        \begin{tikzpicture}
        \begin{axis}[
            width={\textwidth},
            ybar stacked,
            xlabel={\# GPUs},
            ylabel={Wall time (secs)},
            xtick={1,2,3,4},
            xticklabels={1,8,27,64},
        ]
        
        \addplot coordinates {
            (1,0.017966)(2,4.189370)(3,11.091614)(4,13.029241)
        };
        
        \addplot coordinates {
            (1,42.423836-0.017966)(2,46.266260-4.189370)(3,53.186267-11.091614)(4,54.918442-13.029241)
        };
        
        \end{axis}
        \end{tikzpicture}
    \caption{$160 \times 160 \times 160$ grid per GPU}
    \end{subfigure}
        
    \caption{Ratio of communication to computation cost.}
    \label{fig:exp-cost-ratio}
\end{figure}

Figure~\ref{fig:exp-strong-scaling} presents our strong-scaling results.
We fix the number of grid points to $256^3$ and compute the speedup over up to 512 GPUs.
Since a grid with $256^3$ points cannot be loaded on a single GPU because of its memory limit, the $x$-axis starts from 8 GPUs so that each GPU is assigned to $128^3$ grid points.
Table~\ref{tbl:exp-grid-shape} presents the grid shape per GPU we used for decomposing $256^3$ grid for our strong-scaling experiments.

From the results, we observe that our GPU implementation scales well with the number of GPUs. However, we observe that the increase rate of the speedup diminishes as the number of GPUs increases. The reason for this decrease is that the MPI communication cost becomes a dominating factor to the overall cost.

The wall time for 8 GPUs was 24.62 secs, and it took about 3.14 secs for the MPI communication.
In this case, the communication to computation percentage ratio was 12\%$(=3.14/24.62 \times 100)$.
 For 64 GPUs, however, the ratio became 45\% $(=2.40/5.24 \times 100)$, and it was 58\% $(=1.11/1.90 \times 100)$ for 512 GPUs.
Since our implementation on GPUs is significantly faster than the CPU-based implementation, the impact of the communication cost quickly becomes substantial.
We note that the diminishing rate seemed to be accelerated when we started to employ $32^3$ grids (for 128--512 GPUs).
We think this is partly because GPUs were not fully utilized for such small grids, as we discussed in Section~\ref{subsec:exp-one-gpu}.

For weak-scaling tests, we used three different grid schemes per GPU: $64 \times 64 \times 64$, $128 \times 128 \times 128$, and $160 \times 160 \times 160$.
The size of the original grid increases proportionally to the number of GPUs:
for 64 GPUs with $160 \times 160 \times 160$ ($128 \times 128\times 128$ and $64 \times 64 \times 64$) grid per GPU, the shape of the original grid is $640 \times 640 \times 640$ ($512 \times 512 \times 512$ and $256 \times 256 \times 256$); hence, 4 GPUs are assigned to each dimension.

Figure~\ref{fig:exp-weak-scaling} presents our weak-scaling results in terms of efficiency.
A similar trend of diminishing efficiency for larger numbers of GPUs has been observed, as with our strong-scaling results.
The diminishing rate largely depends on the ratio of communication to computation cost.
As we see in Figure~\ref{fig:exp-cost-ratio}, the ratio is higher for $64^3$ grid, 43\% for 64 GPUs, than those for $128^3$ and $160^3$ grids, 27\% and 23\% for 64 GPUs, respectively.
Therefore, the $64^3$ grid showed a worse efficiency than those of the $128^3$ and $168^3$ grids.
In Figure~\ref{fig:exp-cost-ratio}, ``Comp" denotes the pure computation time on each GPU.
In all cases, the computation time on individual GPUs was almost the same.
\section{HIT Decay Simulation}
\label{sec:results}

\begin{table}[t]
\centering
    \caption{Simulation resources used and wall times per time step (seconds) for GPU-enabled and CPU-only simulations of HIT decay on grids with $512^3$ and $1024^3$ points}
    \label{tbl:sim_resources}
    \begin{tabular}{|p{0.1\linewidth}|p{0.1\linewidth}|p{0.23\linewidth} p{0.07\linewidth}|p{0.23\linewidth} p{0.07\linewidth}|}
    \hline
        {Grid} & {\#~nodes} & \multicolumn{2}{c|}{GPU--enabled simulation} & \multicolumn{2}{c|}{MPI--only simulation}\\
        & & {Number of GPUs} & {Wall time} & {Number of MPI ranks} & {Wall time} \\\hline
        $512^3$  & $16$  &   $64$ ($4\times4\times4$) & $0.42$ &  $512$    ($8\times8\times8$) & $5.5$ \\
                 & $128$ &  $512$ ($8\times8\times8$) & $0.083$ & $4096$ ($16\times16\times16$) & $0.72$ \\
        \hline
        $1024^3$ & $128$ &  $512$ ( $8\times8\times8$) & $0.42$ & $4096$ ($16\times16\times16$)& $5.5$ \\
                 & $256$ & $1024$ ($16\times8\times8$) & $0.23$ & $8192$ ($32\times16\times16$) & $2.8$ \\
        \hline
    \end{tabular}
\end{table}

\begin{figure}
    \centering
    \begin{subfigure}{0.49\textwidth}
        \includegraphics[width=\textwidth]{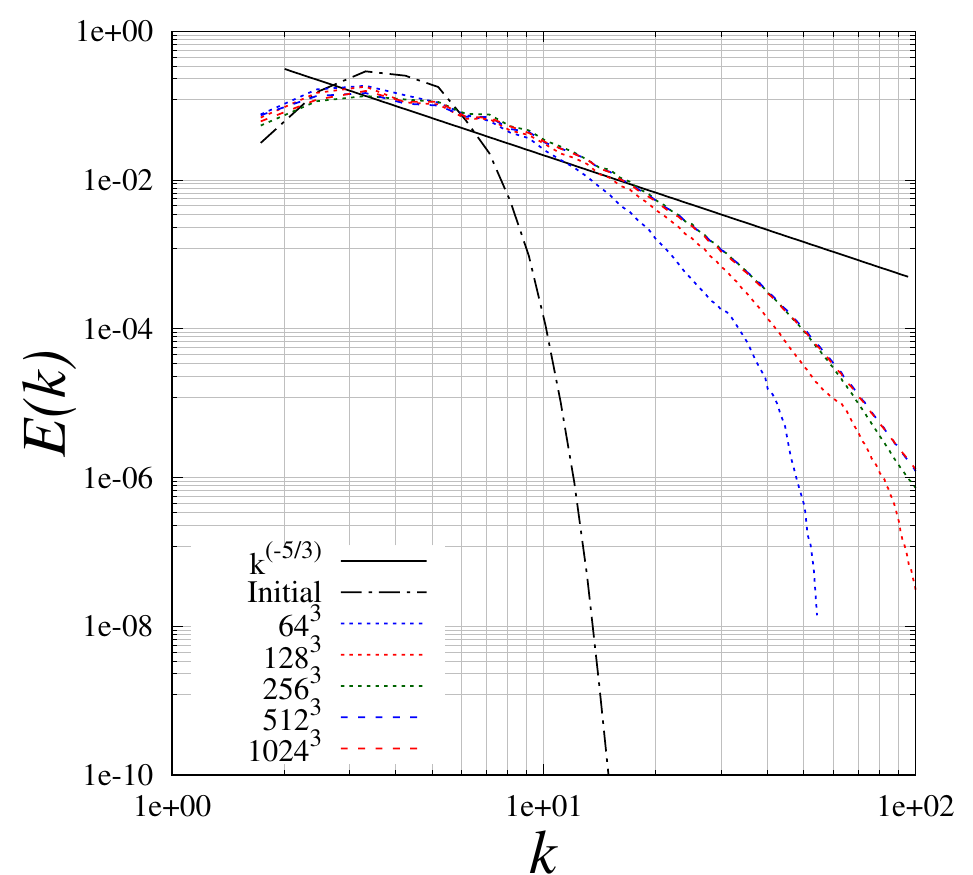}
        \caption{Energy spectrum for HIT decay at $t/\tau=3$ for various grid sizes.}
        \label{fig:spectrum_grid}
    \end{subfigure}
    \begin{subfigure}{0.49\textwidth}
        \includegraphics[width=\textwidth]{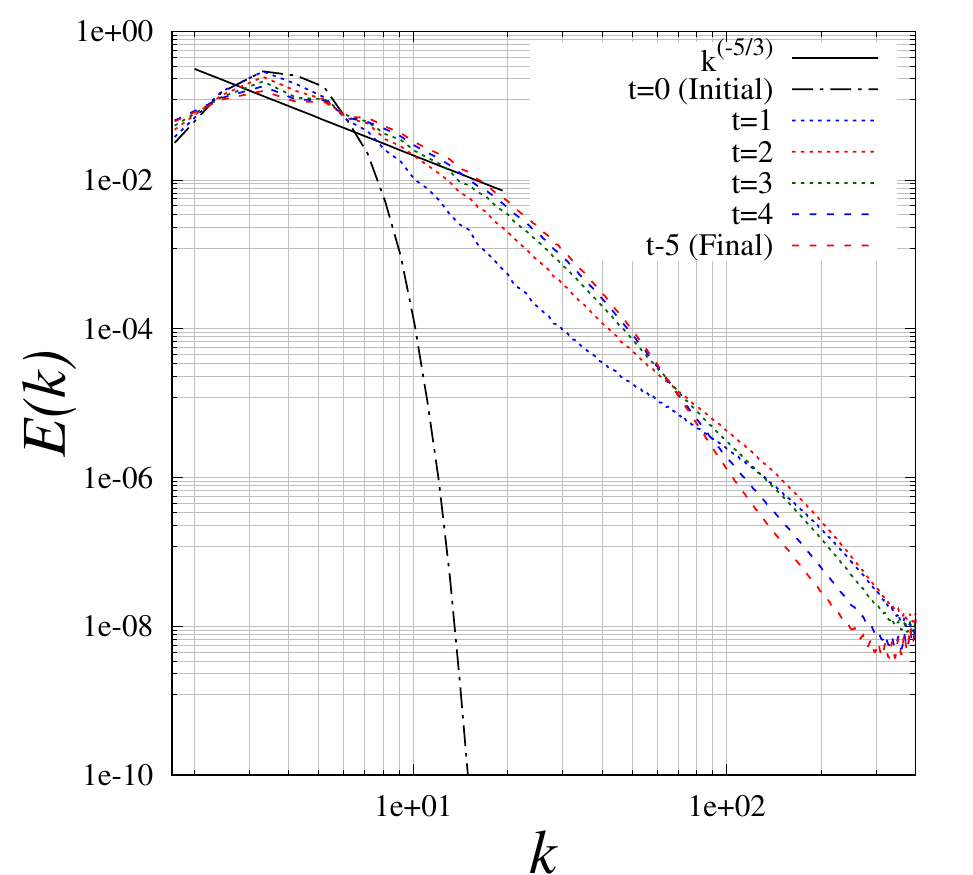}
        \caption{Evolution of the energy spectrum on a grid with $1024^3$ points.}
        \label{fig:spectrum_time}
    \end{subfigure}
    \caption{Energy spectrum for the HIT decay simulations.}
    \label{fig:hit_energy_spectrum}
\end{figure}

\begin{figure}
    \centering
    \begin{subfigure}{0.49\textwidth}
        \includegraphics[width=\textwidth]{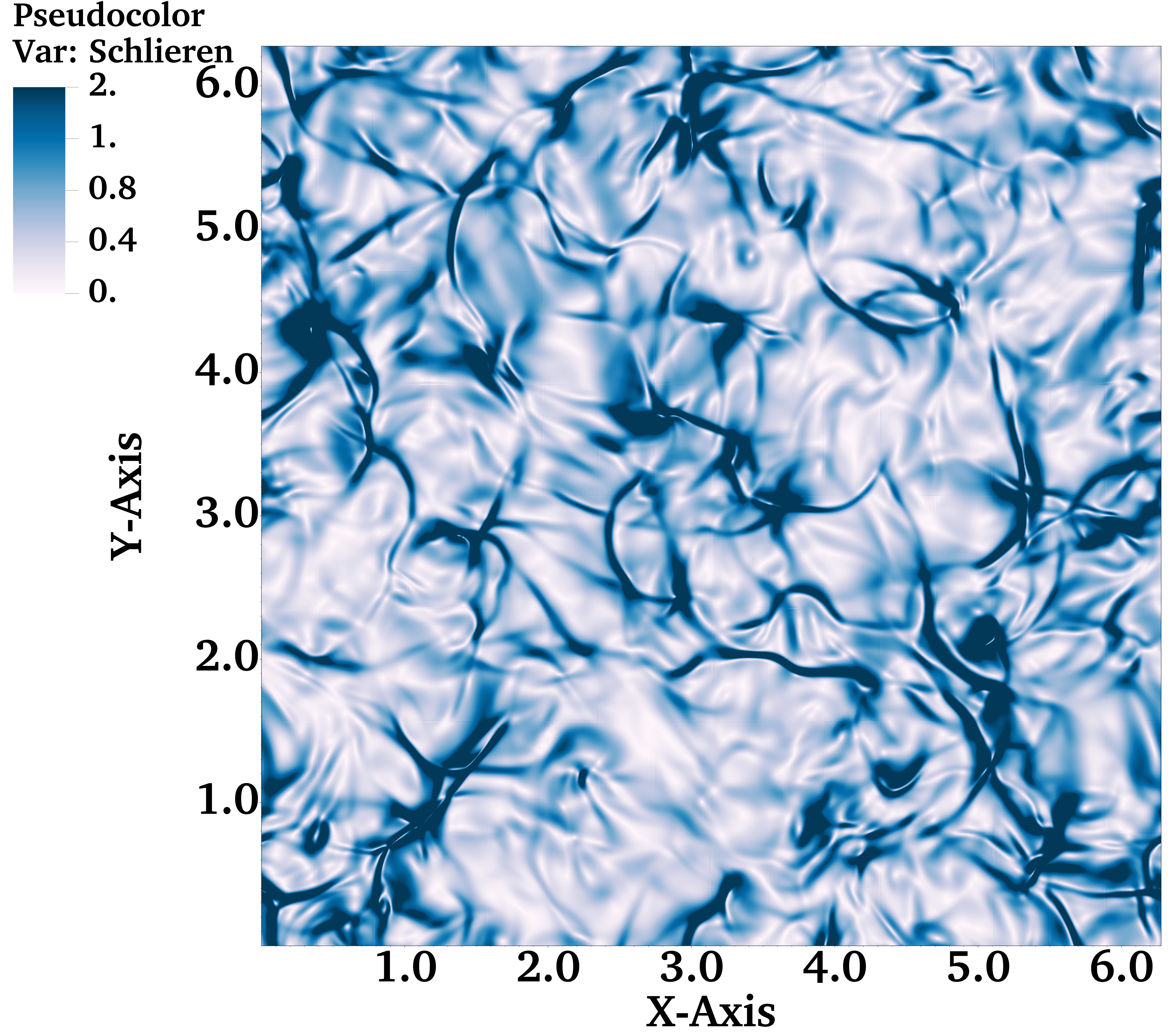}
        \caption{Schlieren ($\|\nabla\rho\|_2$) at $z = \pi$ and $t/\tau=3$.}
        \label{fig:grid1024_schlieren}
    \end{subfigure}
    \begin{subfigure}{0.49\textwidth}
        \includegraphics[width=\textwidth]{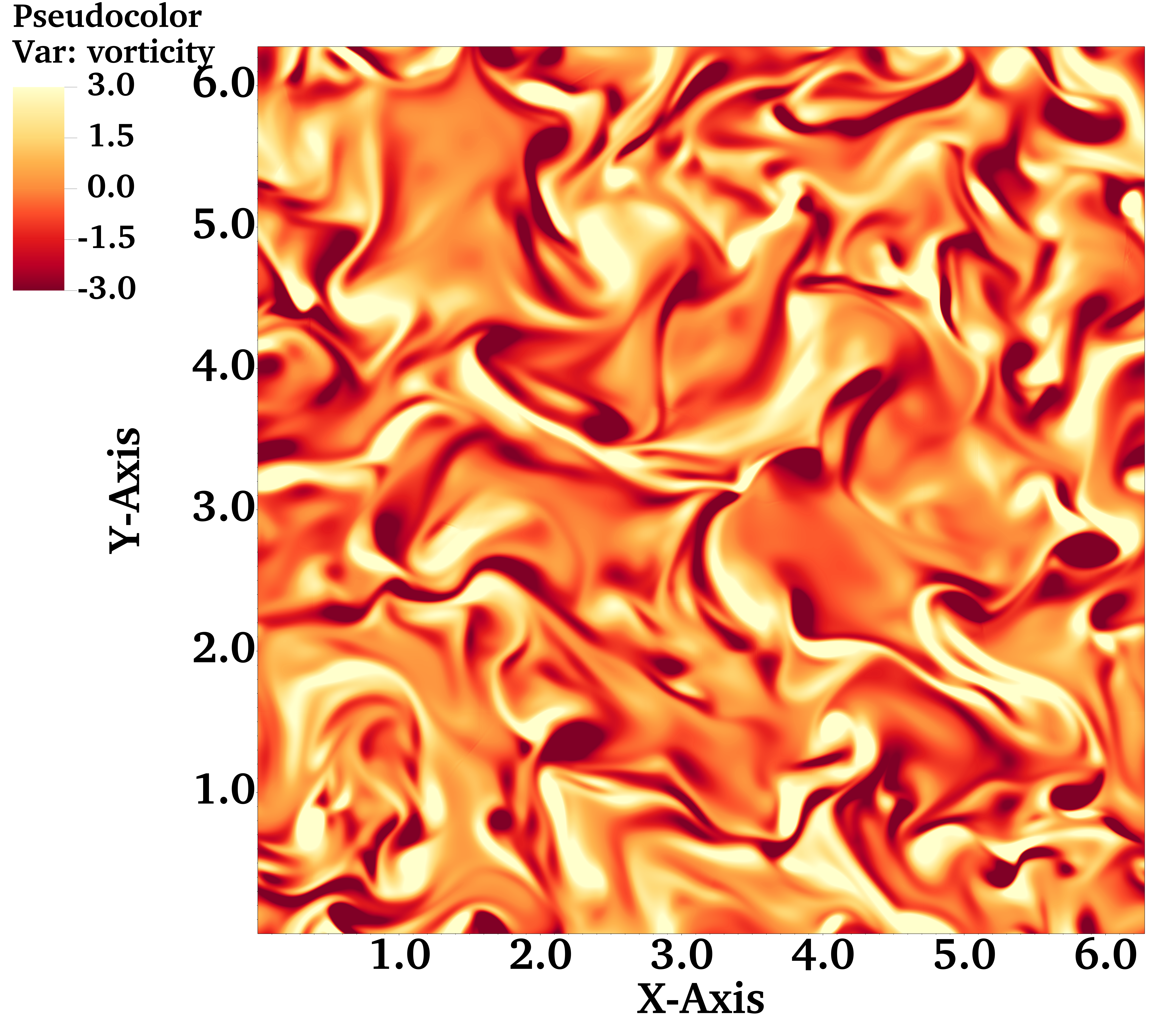}
        \caption{Vorticity magnitude at $z$/=0 and $t/\tau = 3$.}
        \label{fig:grid1024_vorticity}
    \end{subfigure}
    \caption{Isotropic turbulence decay: Solution in the $x$--$y$ plane at $z=0$ on a grid with $1024^3$ points.}
    \label{fig:grid1024_solution}
\end{figure}

In this section we simulate the decay of isotropic turbulence on two large grids: $512^3$ points ($6.7\times10^8$ degrees of freedom) and $1024^3$ ($5.3\times10^9$) points. The simulation setup is the same as described in Section \ref{subsec:isotropic}.\footnote{This specific example is available in the \hypar{}  repository~\cite{hypar} in the subdirectory \tt{Examples/3D/NavierStokes3D/DNS\_IsotropicTurbulenceDecay\_CUDA}.} The results presented here are obtained on LLNL/Lassen; each node has $4$ NVIDIA Volta V100 GPUs and $2$ IBM Power9 CPUs with $40$ cores available for computations, where each core is capable of $4$ hardware threads. The computational performances of GPU-enabled simulations and MPI-only simulations are compared by maximizing the parallelism on the same number of compute nodes. The $512^3$ grid is simulated on $16$ and $1,28$ nodes; thus GPU-enabled simulations use $64$ and $512$ GPUs, respectively, and MPI-only simulations use $512$ and $4096$ MPI ranks, respectively. The $1024^3$ grid is simulated on $128$ and $2,56$ nodes; GPU-enabled simulations use $512$ and $1,024$ GPUs, respectively, and MPI-only simulations use $4,096$ and $8,192$ MPI ranks, respectively. Even though each CPU core is capable of $4$ hardware threads, the MPI-only simulations have only $1$ on a core; we observed performance degradation with $2$ or $4$ MPI ranks per core.

Table~\ref{tbl:sim_resources} summarizes the resources used for these simulations and the observed wall times per time step for the $3$-stage total-variation-diminishing (TVD) Runge--Kutta method in seconds (this does not include the time for file I/O). Overall, we see an excellent speedup for GPU-enabled simulations. While the $512^3$ grid simulation on $1,28$ nodes is around $8$ times faster with the GPU, all the other cases show that the GPU code is $12$--$13$ times faster than the MPI code. We note  that for GPU-enabled simulations, the CPU cores do not do any meaningful computations other than the initial transfer of the solution to the GPU and the transfer of the solution back to the CPU for writing to disk when required. The MPI-only simulations do not use the GPUs at all. Thus, both these approaches do not use the compute nodes to their maximum potential.

Figure~\ref{fig:spectrum_grid} shows the kinetic energy spectra at the final time $t/\tau=3$ for solutions obtained on with $64^3$, $128^3$, $256^3$, $512^3$, and $1024^3$ points. The time step corresponds to a CFL of around $0.4$ for all the simulations. We verified that the difference between the solutions obtained with the GPU-enabled code and the MPI code was within machine roundoff ($10^{-15}$). The energy spectra show that the finer grids are able to resolve the higher wavenumbers (smaller length scales), as expected. Figure~\ref{fig:spectrum_time} shows the time evolution of the energy spectrum for the solution on the $1024^3$-points grid; as the flow evolves, kinetic energy is transferred to the smaller length scales. This underscores the need for efficient scale-resolving simulations. Figures~\ref{fig:grid1024_schlieren} and~\ref{fig:grid1024_vorticity} show the Schlieren and vorticity magnitude plots for the solution obtained on the grid with $1024^3$ points; the flow is dominated with small-scale structures as well as localized regions of very high gradients.
\section{Conclusions}
\label{sec:conclusion}

In this work we explore the development of \hypar~\cite{hypar}, a CPU-based C/C++ high-performance code, to run efficiently on GPU-based computing platforms. The naive use of the SIMD directives in an existing CPU codebase, by primarily targeting memory operations across the CPU-GPU bus, leads to severe performance degradation on GPUs because GPUs require a fundamentally different programming paradigm from that of  CPUs. A bottom-up approach that focused on forming the equations directly on GPUs significantly reduces traffic between CPUs and GPUs, across the PCIe bus/NVLink, and improves the performance of the code  significantly.

This work focuses on the three-dimensional compressible Navier--Stokes equations discretized by a nonlinear fifth-order WENO scheme in space and an explicit Runge--Kutta method in time. This combination of numerical discretization schemes allows us to exploit high potential parallelism and scalability while maintaining high numerical accuracy and spectral resolution. We demonstrate our results on a canonical turbulent flow problem, the decay of isotropic turbulence in a periodic domain characterized by the transfer of energy from larger to smaller length scales. We conduct DNS studies by resolving the Kolmogorov scale on a grid of size $1024^3$ ($\sim 5$ billion degrees of freedom).

The implementation of a high-order shock-capturing algorithm in this paper is among the earliest for simulating HIT in compressible flows on heterogeneous platforms. We carry out simulations on up to 1,024 GPUs with CUDA-aware MPI communication across the nodes. Our results demonstrate that the optimizations proposed in this paper lead to more than 200x reduction in computation time using multiple threads on a GPU compared with a single CPU core. The scalability of the GPU-accelerated code is evaluated by using strong and weak scaling, and we investigate the impact of communication cost on the overall computation time. We compare our GPU-enabled and MPI-only implementations on DOE leadership-class supercomputers with multicore CPUs and GPUs, and we observe that the GPU-enabled code performs around $12$--$13$ times faster on the these architectures. The strategies presented in this paper extend to conventional shock-capturing schemes, including low-dissipation, bandwidth-optimized WENO schemes that are better suited to DNS studies. In ongoing work, we are exploring nonlinear compact schemes, such as the CRWENO or hybrid compact-WENO schemes with higher spectral resolution. However, they require the direct solution of tridiagonal systems of equations; this presents additional challenges for GPUs.



\bibliographystyle{elsarticle-num}
\bibliography{citation}

\begin{center}
	\scriptsize \framebox{\parbox{4in}{Government License (will be removed at publication):
			The submitted manuscript has been created by UChicago Argonne, LLC,
			Operator of Argonne National Laboratory (``Argonne").  Argonne, a
			U.S. Department of Energy Office of Science laboratory, is operated
			under Contract No. DE-AC02-06CH11357.  The U.S. Government retains for
			itself, and others acting on its behalf, a paid-up nonexclusive,
			irrevocable worldwide license in said article to reproduce, prepare
			derivative works, distribute copies to the public, and perform
			publicly and display publicly, by or on behalf of the Government.
			The Department of Energy will provide public access to these results of federally sponsored research in accordance with the DOE Public Access Plan.
			http://energy.gov/downloads/doe-public-access-plan.
			\\
			\\
			LLNL-JRNL-836899
}}
	\normalsize
\end{center}





\end{document}